\documentclass[11pt]{article} 

\usepackage{graphicx,amstext,amssymb}

\begin{document}

\author{S.V. Akkelin$^{1}$, Yu.M. Sinyukov$^{1}$}
\title{Phase-space densities and effects of resonance decays in
 hydrodynamic approach to heavy ion collisions }
\maketitle

\begin{abstract}
A method allowing  analysis of the overpopulation of phase-space
in heavy ion collisions in a model independent way is proposed
within  the hydrodynamic approach. It makes it possible to extract
a chemical potential of thermal pions at freeze out irrespective
of the form of freeze-out (isothermal) hypersurface in Minkowski
space and transverse flows on it. The contributions of resonance
(with masses up to 2 GeV) decays to spectra, interferometry
volumes and phase-space densities are calculated and discussed in
detail. The estimates of average phase-space densities and
chemical potentials of  thermal pions are obtained for  SPS and
RHIC energies. They demonstrate that multibosonic phenomena at
those energies might be considered  as a correction factor rather
than as a significant physical effect. The analysis of the
evolution of the pion average phase-space density in chemically
frozen hadron systems shows that it is almost constant or slightly
increases with time while the particle density and phase-space
density at each space point drops down rapidly during the system's
expansion. We found that, unlike the particle density, the average
phase-space density has no direct link to the freeze-out criterion
and final thermodynamic parameters, being connected rather to the
initial phase-space density of hadronic matter formed in
relativistic nucleus-nucleus collisions.
\end{abstract}
\begin{center}
{\small {\it $^{1}$ Bogolyubov Institute for Theoretical Physics,
Kiev 03143, Metrologichna 14b, Ukraine. \\[0pt] }}

PACS: {\small {\it 24.10.Nz, 24.10.Pa, 25.75.-q, 25.75.Gz,
25.75.Ld.}}

 Keywords: {\small {\it relativistic heavy ion collisions,
phase-space density, hydrodynamic evolution, hadron resonance gas,
inclusive spectra, HBT correlations.}}

 Corresponding author: {\small {\it Yu.M. Sinyukov, Bogolyubov
Institute for Theoretical Physics, Kiev 03143, Metrologichna 14b,
Ukraine. E-mail: sinyukov@bitp.kiev.ua, tel. +380-44-2273189,
tel./fax: 380-44-4339492}}
\end{center}

\section{Introduction}
The interferometry method to measure particle phase space
densities averaged over the volume of a system that emits the
particles was first proposed by Bertsch \cite{Bertsch}. If systems
formed in ultrarelativistic A+A collisions are thermal, then it is
possible to compare the experimental value of the average
phase-space density (APSD) with one associated with the
Bose-Einstein \cite{Ferenc} or Fermi-Dirac \cite{Xu} thermal
distributions, $ f_{BE,FD}(p),$ and extract in this way a particle
chemical potential (or fugacity) at the freeze-out temperature.
Such a chemical potential could have a positive value even for
pions since a hadron gas evolution is  chemically frozen rather
than chemically equilibrated
\cite{Koch,Teaney,Braun-Munzinger,Hirano,A-B-S}. If it  turns out
that the particle fugacity is fairly large, then the boson density
could be around the critical one and significant multibosonic
\cite{Ledn-Sin,Akkelin,Heinz,QHZ}, and coherent \cite{A-L-S}
effects might be present at low particle momenta.

There are, however, two serious problems which have to be solved
to apply the above method. It is obvious that the APSD
$\left\langle f(p)\right\rangle $ can be associated directly with
a thermally equilibrated  distribution for  static homogeneous
fireballs, $f_{BE}(p)$ for bosons. For expanding inhomogeneous
systems, characterized by a set of (local) equilibrium
distributions which are Doppler shifted in the vicinities of each
space-time point according to the collective velocity field, the
phase-space density \textit{averaged over some hypersurface}
situated at or "after" the freeze-out could lose, in general, the
direct connection to the Bose-Einstein or Fermi-Dirac phase-space
distributions. As it was demonstrated in Ref. \cite{Tomasik}, even
at a uniform temperature and chemical potential at  freeze-out,
the intensity and profile of longitudinal and transverse flows
influence very essentially the value of $\left\langle
f(p)\right\rangle $, making, thus, the extraction of fugacity by
comparison with  $f_{BE}(p)$  ambiguous and model dependent. One
of the aims of this paper is to study and substantiate the model
independent possibility for an analysis of overpopulation of the
phase-space in 3D expanding fireballs at  freeze-out.

The another problem is resonance contributions to particle spectra
and interferometry radii \cite{Padula}. If chemical freeze-out
takes place, then about 2/3 of pions comes from decays of
resonances in high energy heavy ion collisions
\cite{Braun-Munzinger,A-B-S}. Only a small part of them are
long-lived resonances which give the effect of a halo
\cite{Csorgo} - a suppression of correlation functions of
identical pions (see also \cite{A-L-S}). The others, short-lived,
enhance significantly the pion single-particle spectrum, as
compare to the thermal pion one, and also modify the
interferometry radii. Our analysis of pion production by some
\textit{concrete} resonance species shows that even for thermal
distribution of those resonance species at freeze-out the spectrum
of the produced pions is far from the thermal one. Also, an
effective region of such a pion emission at small momenta is not
associated with the interferometry volume at the corresponding
small mean momenta of pion pairs from the resonance decays.
Peculiarities of pion production by resonances are important to
understand the influence of resonance decays on the average
phase-space densities which are currently measured in experiments
at CERN SPS and BNL RHIC. We analyse quantitatively the total
effect on spectra and interferometry volume of pion production by
huge number of resonances with masses up to 2 GeV and dependence
of the effect on the intensity of flows. As a result, we propose
how to cope with the resonance contributions to study the
freeze-out properties of the expanding systems.

One of the most serious challenges to the theoretical
understanding of the processes of A+A collisions is multiplicity
(and energy) dependence of the APSD $\left\langle f\right\rangle $
in central events. According to the Bertsch result \cite{Bertsch}
$\left\langle f\right\rangle \sim$
$\frac{dN/dy}{T_{eff}^{3}V_{int}}.$ If at different energies of
A+A collisions the pion interferometry volume $V_{int}$ of the
created systems is proportional to the pion multiplicity $dN/dy$,
and $T_{eff}$, the effective temperature of the transverse spectra
of pions, changes only slightly, then approximate constancy of the
APSD could be considered a the natural property of the system at
freeze-out in high energy A+A collisions. It was a very likely
hypothesis at relatively small energies \cite{Ferenc}, since it
had been supported by the observed proportionally between $dN/dy$
and $V_{int}$ \cite{NA35}. However, this proportionally law starts
to be violated already at  SPS energies and is completely
destroyed  in experiments at RHIC \cite{Adamova}. In the latter
case the interferometry radii and volume turn out to be
essentially smaller than was expected at the corresponding RHIC
multiplicities (the HBT puzzle) \cite{Pratt}. As a consequence,
the APSD at RHIC energies is found to be the considerably higher
than at the formerly achieved energies \cite{Ray}. Does this mean
that freeze-out conditions of the system  in heavy ion collisions
at RHIC are essentially different from the ones at SPS? Does this
signify that the freeze-out particle densities at RHIC come close
to the critical value? The latter could lead to the multibosonic
effects at freeze-out which, in  turn, can reduce the observed HBT
radii \cite{Ledn-Sin,Akkelin}. Are thus multibosonic effects
relevant to the HBT puzzle?

To understand the above problems we study the APSD and particle
density temporal evolution in a few hydrodynamic models. We
observe,  in particular, that the phase-space density averaged
over the freeze-out volume, $\left\langle f(p)\right\rangle $,
drops rather slowly during the 3D expansion of the system, - much
slower than the phase-space density $f(t,\mathbf{r},\mathbf{p} )$
at any fixed $\mathbf{r}$ and than the particle density. This
difference in the behavior of the two groups of values becomes
bigger when the intensity of the transverse flow grows. The
detailed analysis of this effect as well as the model independent
estimates of chemical potential for the thermal pions at
freeze-out makes it possible to shed light on the reason of the
relatively high APSD at  RHIC and on the HBT puzzle.

\section{Evolution of the average phase-space density.}
\subsection{Temporal behaviour of APSD $\langle f(p)\rangle$ in
analytically solved hydrodynamic models} Let us demonstrate the
important features of the APSD evolution versus particle density
one by one using hydrodynamic models of A+A collisions. To see the
main idea we start from the exact solutions of perfect
non-relativistic hydrodynamics. If the initial conditions
correspond to the thermal Boltzmann distribution with uniform
temperature $T_{0}$, zero initial flows, $\mathbf{u}(t=0,
\mathbf{r})=0$, and spherically symmetric Gaussian (with radius
$R_{0}$) profile of particle density, then the hydrodynamic
evolution of that system is described by the local equilibrium
phase-space distribution which is an exact solution of Boltzmann
equation for any cross-section of particle interactions
\cite{Csorgo1} and has the form:
\begin{equation}
f^{l.eq.}(t,\mathbf{r},\mathbf{p})=\frac{N}{R_{0}^{3}}\left( \frac{1}{(2\pi
)^{2}mT_{0}}\right) ^{\frac{3}{2}}\!\exp (-\frac{m(\mathbf{v}-\mathbf{u}(%
\mathbf{r}))^{2}}{2T(t)}+\!\frac{\mu (r)}{T(t)}),  \label{hydro}
\end{equation}
where $N$ is total number of particles,$\ \mathbf{v}=\mathbf{p}/m$ are their
velocities, and $T(t)=\frac{T_{0}}{1+T_{0}t^{2}/mR_{0}^{2}},\frac{\mu (r)}{%
T(t)}=-\frac{\mathbf{r}^{2}}{2R_{0}^{2}}\frac{mR_{0}^{2}}{T_{0}t^{2}}/(1+\frac{%
mR_{0}^{2}}{T_{0}t^{2}}),\mathbf{u}(\mathbf{r})=\mathbf{r}\frac{tT(t)}{mR_{0}^{2}%
}$.

One can check directly that the momentum spectrum and the
interferometry radii, evaluated at any time $t$ (at any isotherm)
are identical to those calculated at the initial time $t=0$. The
easiest way to see it is to rewrite  expression (\ref{hydro}) in
the form

\begin{equation}
f^{l.eq.}(t,\mathbf{r},\mathbf{p})=\frac{N}{R_{0}^{3}}\left( \frac{1}{(2\pi
)^{2}mT_{0}}\right) ^{\frac{3}{2}}\!\exp (-\frac{\mathbf{p}^{2}}{2mT_{eff}}%
-\!\frac{(\mathbf{r-}t\,\mathbf{p}/m)^{2}}{2R_{0}^{2}}),  \label{free}
\end{equation}
where the effective temperature of the spectra is a constant in
time and {\it{equal to the initial one}}, $T_{eff}=T_{0}$. Thus,
the maximal phase space density and Bertsch's average phase-space
density \cite{Bertsch}

\begin{equation}
\left\langle f(t,\mathbf{p})\right\rangle =\frac{\int \left(
f^{l.eq.}(t,\mathbf{r},\mathbf{p})\right) ^{2}d^{3}r}{\int f^{l.eq.}(t,%
\mathbf{r},p)d^{3}r}\mathbf{=}\ \frac{N}{R_{0}^{3}}\left( \frac{m}{2(2\pi
)^{2}T_{0}}\right) ^{\frac{3}{2}}\!\exp (-\frac{\mathbf{p}^{2}}{2mT_{eff}})
\label{Bertsch}
\end{equation}
is the same at each $\mathbf{p}$ as it was at the initial time!

The space density of the particles falls down with time as
$1/t^{3}$ at large $t$. So, the particles in the system will
really stop  interacting at some later stage and the system
decouples, preserving its maximal PSD and  average PSD values as
they were at the moment when the fireball was created. It is
noteworthy that the local phase-space density $f^{l.eq.}(t,\mathbf{r},%
\mathbf{p})$ in the system drops with time at any fixed point $\mathbf{r\ }$%
very rapidly, as $\exp [-(c_{1}^{2}t^{2}+c_{2}t)]$.

The perfect constancy of the APSD is caused by the peculiarities
of this specific model: in particular, its spherical symmetry, as
 was discussed in detail in Ref. \cite{Sinyukov}. At the same
time, as we will demonstrate, the important feature of this simple
model, namely, the principal distinction in temporal behaviours of
different types of densities, is preserved in realistic models of
A+A collisions: the APSD decreases with time essentially slower
than the particle density.

To make the next step towards  clarifying the effect, let us
consider one more analytical illustration of it based on the model
of a 3D expanding fireball with longitudinally Bjorken-like flow
$v_{z}=r_{z}/t$ . Here is the exact non-relativistic solution of
the ideal hydro equations for an initially Gaussian transverse
density profile which results in the following evolution of the
locally equilibriated  phase-space density \cite{A-B-S}:

\begin{eqnarray}
f^{l.eq.}(t,\mathbf{r},\mathbf{p}) &=&\left( \frac{1}{2\pi mT}\right)
^{3/2}n\exp (-\frac{m}{2T}(\mathbf{v}-\mathbf{u)}^{2}),  \nonumber \\
n(t,\mathbf{r}) &=&n_{0}\frac{R_{0}^{2}t_{0}}{R^{2}(t)t}\exp (-\frac{%
r_{x}^{2}}{2R^{2}}-\frac{r_{y}^{2}}{2R^{2}}),  \nonumber \\
\mathbf{v} &=&\frac{\mathbf{p}}{m}, \mathbf{u}(t,\mathbf{r})=\left( \frac{%
\stackrel{\cdot }{R}}{R}r_{x},\frac{\stackrel{\cdot }{R}}{R}r_{y},\frac{1}{t}%
r_{z}\right),  \label{F-def}
\end{eqnarray}
where the initial particle density in the central part of system,
$r_{x}=r_{y}=0$, is $n_{0}=N/(2\pi )^{3/2}R_{0}^{2}t_{0}$ and
\begin{equation}
T=T_{0}\left( \frac{R_{0}^{2}t_{0}}{R^{2}(t)t}\right) ^{2/3},R\stackrel{%
\cdot \cdot }{R}=\frac{T}{m}.  \label{hyd-sol2}
\end{equation}
At each moment $t$ the longitudinal spectrum in this model is $m\frac{dN}{%
dp_{z}}=$ $2\pi R_{0}^{2}t_{0}n_{0}=const$, similar to $\frac{dN}{dy}=$ $%
const$ in the relativistic boost-invariant model, and the
momentum spectrum is

\begin{equation}
\frac{d^{3}N}{d^{3}p}=\frac{n_{0}R_{0}^{2}t_{0}}{m^{2}T_{eff}(t)}\exp (-%
\frac{p_{T}^{2}}{2mT_{eff}(t)}),  \label{spect-sol2}
\end{equation}
where the effective temperature, $T_{eff}$, is \
\begin{equation}
T_{eff}=m\stackrel{\cdot }{R}^{2}+T.  \label{T_eff}
\end{equation}
The $T_{eff}(t)$ does not change much since the decrease of
temperature $T$ when the system expands is accompanied by the
increase of transverse flows described by $\stackrel{\cdot }{R}$
in Eq. (\ref{hyd-sol2}). The typical behavior of the temperature
and effective temperature in this model  can be seen in Fig. 1.

The  interferometry radii, \textit{out-, side-} and \textit{long-
} take the forms
\begin{equation}
R_{O}^{2}=R_{S}^{2}=\frac{T}{T_{eff}}R^{2},\;R_{L}^{2}=t^{2}\frac{T\,}{m}
\label{int-radii}
\end{equation}
 and are similar to the ones in the  relativistic case at the central rapidity point
  \cite{Csorgo2,AkkSin}, if $m\rightarrow m_{T}\equiv
\sqrt{m^{2}+p_{T}^{2}}$. The interferometry volume, taking into
account (\ref {hyd-sol2}),  is
\begin{equation}
V_{int}\equiv R_{O}R_{S}R_{L}=\frac{R_{0}^{2}t_{0}(T_{0})^{3/2}}{T_{eff}%
\sqrt{m}}  \label{vol1}
\end{equation}
and does not change significantly since it depends on $T_{eff%
}$ only. So, it is not surprising after all, that the APSD is, at
least approximately, preserved with time

\begin{equation}
\left\langle f(t,\mathbf{p})\right\rangle =n_{0}\left( \frac{1}{4\pi mT_{0}}%
\right) ^{3/2}\exp (-\frac{p_{T}^{2}}{2m}\frac{1}{T_{eff}(t)}).  \label{PSDA}
\end{equation}
The average momentum phase-space density, $\left\langle f(t,\mathbf{%
p})\right\rangle $, as well as the interferometry volume
$V_{int}$,
will be preserved in time if and when, at some stage of the evolution, $%
T_{eff}(t)$ is constant in time.

Practically, the effective temperature, and thus the APSD $\left\langle f(t,%
\mathbf{p})\right\rangle $, could be slightly reduced or even
increased with time during the evolution, depending on initial
conditions, the evolution stage and particle mass. Behavior of the
effective temperature is correlated unambiguously with the
behavior of  particle transverse energy per unit of rapidity (or
average particle transverse momentum). In the idealized case of
the 1D longitudinal expansion of a one-component gas, the particle
transverse energy per unit of rapidity can  only decrease since
the effective temperature of the transverse spectrum in that case
is just the system's temperature that falls down permanently.
However, in a realistic case when transverse expansion is switched
on, the  transformation of a system's heat energy into the
transverse and longitudinal kinetic energies of particles becomes
much more complicated. In the 3D expansion the diminution of the
total energy of the part of the hydrodynamic tube confined within
a small rapidity interval is accompanied by a faster decrease in
the particle longitudinal kinetic energy, comparing to the 1D
case. It is caused by faster temperature decrease, while
longitudinal collective velocities are the same in that piece of
fluid. As for the transverse energy, it  accounts for not only a
decrease of the heat energy, associated with temperature, but also
an increase with time of transverse flows. Therefore, the particle
transverse energy (in the relativistic case it is better to
consider average transverse momentum of particles) essentially
does not change during the evolution. Such a process of
transformation of  heat into  transverse flow is accompanied by an
appearance of positive correlations between the absolute value of
particle transverse momentum and its distance from the centre of
the system. The degree of that correlation grows with time
compensating for the contribution of increasing geometrical radii
to the interferometry volume $V_{int}(p_{T})$ \cite{AkkSin}. Since
the transverse momentum
spectra do not change much, the APSD $\left\langle f(t,\mathbf{p}%
)\right\rangle \sim \frac{d^{3}N}{dyd^{2}p_{T}}/V_{int}(p_{T})$
becomes about
constant at some stage of evolution unlike the local phase-space density $f(t,%
\mathbf{r,p})$ and particle density. The latter is just a fast
monotonically decreasing function (see (\ref{F-def})):

\begin{equation}  \label{density}
n(t,\mathbf{0})=n_0\frac{R_0^2t_0}{R^2(t)t}\stackrel{t\rightarrow \infty }{%
\longrightarrow }\frac 1{t^3}
\end{equation}
This is one of our main points - the system can reach small
densities and decay with rather high APSD reflecting the high
average phase-space density at the moment when the system was
formed!

Now we pass from general physical ideas to the concrete results
within the
realistic model of Pb+Pb collisions at CERN energies $158$ AGeV \cite{A-B-S}%
. The model describes hydrodynamically the evolution of chemically
frozen hadron-resonance gas from the hadronization stage to the
thermal, or kinetic, freeze-out. As  was found recently
\cite{Teaney}, if chemical freeze-out is incorporated into
hydrodynamics, then the final spectra and fireball lifetimes are
insensitive to the temperature at which the switch from
hydrodynamics to cascade RQMD is made.  This implies that the
local conservation law describing the transformation of heat
energy into collective flows for a chemically frozen hydrodynamic
evolution results in spectra which correspond approximately to the
microscopic cascade calculations. Based on an analysis of the
particle number ratios, pion rapidity density at $y\approx 0,$
slopes of transverse spectra and values of interferometry radii
for pions, kaons and protons, the reconstruction of the
hadronization stage was done in Ref. \cite{A-B-S} by solving
backwards  in time the hydrodynamic equations for the chemically
frozen composition of mesons and baryons with masses up to 2 GeV.
The ''initial conditions'' for such a solution were defined from
an analysis of spectra and correlations at the thermal freeze-out
stage. Supposing that the hadronization of Bjorken-like expanding
quark-gluon plasma (QGP) happens at a uniform chemical freeze-out
temperature $T_{ch}$  with Gaussian-like particle density
transverse
profile of some width $R$ and self-similar transverse flow $y_{T}=r%
\stackrel{\cdot }{R}/R$ , one can restore from the
hydrodynamic solutions the hadronization proper time $%
\tau _{ch}$, transverse radius $R(\tau_{ch})$ , transverse
velocity $v_{R}(\tau _{ch})${$=$}$\stackrel{.}{R}(\tau _{ch})$ at
the Gaussian ''boundary'' of the system, and then the energy and
particle number densities in the center of the system, $\epsilon
_{ch}$ and $n_{ch}$. The temperature $T_{ch}$ and baryonic
chemical potential $\mu _{B}(\tau _{ch})$ are defined in Ref.
\cite{A-B-S} from the analysis of particle number ratios.

\begin{center}

TABLE 1. The characteristics of the chemical freeze-out in Pb+Pb
SPS collisions at 158 AGeV taken from Ref. \cite{A-B-S}.

\bigskip

\begin{tabular}{|c|c|c|c|c|c|c|}
\hline
$T_{ch}$ GeV & $\mu _{B,ch}$ GeV & $\tau _{ch}$ fm & $R_{ch}$ fm & $%
v_{R,ch}$ & $\varepsilon _{ch}$ $\frac{GeV}{fm^{3}}$ & $n_{ch}$ $%
\frac{1}{fm^{3}}$ \\ \hline
$0.164$ & $0.224$ & $7.24$ & $4.63$ & $0.304$ & $0.420$ & $0.421$ \\ \hline
\end{tabular}
\end{center}

The evolution of the APSD $\left\langle f(\tau ,p)\right\rangle $
for the frozen number of direct pions in hadron-resonance gas with
the initial conditions at chemical freeze-out as in Table 1 is
demonstrated in Fig. 2.

 As one can see, the APSD is
changed less than $10$ per cent  during the evolution from the
hadronization stage to the kinetic freeze-out  and its values at
small transverse momenta are even higher at the final than at the
initial moment. At the same time, as is shown in Fig. 3, the
densities of direct pions drop rather quickly in the central part
of the system where soft momentum particles are produced. One can
see the principal difference  in the evolutions of the space- and
average phase-space densities from the comparison of correspondent
curves at $\tau =15$ fm/c and $\tau =7.24$ fm/c. While the APSD
changes not more than $35$ per cent and even increases at small
$p_{T}$, the particle density drops a factor of 6 ! We can
conclude, therefore, that the basic physical phenomenon of
approximate ''conservation'' of the pion APSD during the system
evolution that we demonstrated analytically in the
non-relativistic models of the 3D hydrodynamic evolution also
takes place in the general case of relativistic chemically frozen
3D expansion of a many-component hadron resonance gas.
\subsection{Totally averaged phase-space density $\langle f\rangle$ as integral of motion}
 Let us devote the end of this Section to the new, important
results concerning the evolution of the phase-space density
$\left\langle f(t)\right\rangle$ that is \textit{totally} averaged
over momentum and space with the integral measure $d^{3}rd^{3}p$ :

\begin{equation}
\left\langle f(t)\right\rangle \equiv \frac{ \int f^{2}(t,\mathbf{r}%
,p)d^{3}rd^{3}p}{ \int f(t,\mathbf{r},p)d^{3}rd^{3}p},
\label{APSD-tot}
\end{equation}
It is obvious that $\left\langle f(t)\right\rangle \neq \int
d^{3}p\left\langle f(t,\mathbf{p})\right\rangle .$ Supposing the
Boltzmann non-relativistic approximation for the distribution
function holds:

\begin{equation}
f(t,\mathbf{r},\mathbf{p})=(2\pi )^{-3}\exp \left( \frac{m(\mathbf{v-u(}t%
\mathbf{,r))^{2}}}{2T(t,\mathbf{r})}\right) \xi (t,\mathbf{r})
\label{distrib}
\end{equation}
where $\mathbf{u(}t\mathbf{,r)\ }$ is the collective velocity
field,  $\xi \mathbf{(}t\mathbf{,r)=}\exp (\mu /T)$ is the
fugacity, and $\mu $ is the (non-relativistic) chemical potential,
one can find for the  finite expanding system that
$\frac{d}{dt}\left\langle f(t)\right\rangle =0$.

Indeed, from the non-relativistic hydrodynamic equations for the
one-component Boltzmann gas (see, e.g., \cite {Huang}) it follows
the fugacity $\xi $ is preserved along the current lines:
$(\partial _{t}+u_{i}\partial _{r_{i}})\xi =0$. Then one should
appeal again to the hydro equations to deduce that
\begin{equation}
(\partial _{t}+\partial _{r_{i}}u_{i})T^{3/2}\xi =0,  \label{proof1}
\end{equation}
and get, finally, using the property of finiteness of the system, that ($%
k=1,2,...)$

\begin{eqnarray}
\frac{d}{dt}\left( \int d^{3}r\int f^{k}(t,\mathbf{r},\mathbf{p}%
)d^{3}p\right) &=&\int d^{3}r\frac{\partial }{\partial t}\left( \int f^{k}(t,%
\mathbf{r},\mathbf{p})d^{3}p\right) =  \nonumber \\
-\left( \frac{m}{2\pi k}\right) ^{3/2}\int d^{3}r\partial
_{r_{i}}(u_{i}T^{3/2}\xi ^{k}) &=&0  \label{proof}
\end{eqnarray}
So the totally averaged phase-space density is conserved during
the system expansion! Such  models which describe initially finite
systems, $f(t_{0},r \rightarrow \infty) \rightarrow 0$, belong to
the class of Landau-like models \cite{Landau}. Of course, the
property $\left\langle f(t)\right\rangle =const$ is satisfied in
the concrete model (\ref{hydro}) of spherically expanding
fireball.

There is, however,  another, Bjorken-like class of hydrodynamic
models, which are  boost invariant and, so, formally infinite in
the longitudinal direction. For this situation it is better to
define the phase-space densities  averaged over all phase-space
except particle longitudinal momentum $p_{z}$ , or rapidity $y,$
in order to
associate the APSD with the small rapidity interval near mid-rapidity, $%
v_{L}=p_{z}/m\approx 0$, or $y\approx 0.$ We will mark the
corresponding APSD as $\left\langle f(t,y)\right\rangle .$ Since
the boost-invariance dictates that distribution function depends
just on the difference $v_{L}-u_{L}$ ($u_{L}=r_{z}/t$) for the
non-relativistic case,  or $y-\eta $ ($\eta $ is hydrodynamic
rapidity ) for the relativistic one,  the integration over fluid
longitudinal coordinates at fixed \textit{particle} velocities
$v_{L}\approx 0$, or $y\approx 0$, in (\ref{APSD-tot}) can be
substituted with the integration over particle
rapidity at fixed \textit{hydrodynamic} velocities $u_{L}$ $%
\approx 0$, or $\eta \approx 0.$ Therefore,  $\left\langle
f(t,y)\right\rangle $
corresponds to a totally averaged phase-space density in a small central slice $%
\Delta \eta$ of the hydrodynamic tube that expands conserving
entropy $S$ and particle number $(dN/d\eta)\Delta \eta$. Using Eq.
(\ref{proof1}) one can find, similar to (\ref{proof}), that
\begin{equation}
\frac{d}{dt}\left( \int td^{2}r\int f^{k}(t,\mathbf{r},\mathbf{p}%
)d^{3}p\right) =0,  \label{proof2}
\end{equation}
and, thus,  the APSD $\left\langle f(t,y)\right\rangle$ in this
case is also preserved in time.

It is useful to explain and generalize to the  relativistic case
the above results without resorting to
hydrodynamic equations. Let us consider the time evolution of the set of $i-$%
fluid elements starting from some initial time $t_{0}$. They
contain particle numbers $\Delta N_{i}$, entropy $\Delta S_{i}$,
energy in their rest frames  (marked by asterisk) $\Delta
E_{i(0)}^{\ast }$ and occupy initially the small volumes $\Delta
 V_{i(0)}^{\ast }$ near points $x_{i(0) }$ with temperatures
$T_{i(0)}$ and chemical potentials $\mu_{i(0)}$. At some later
time $t_{f}$ the fluid elements are characterized by the
corresponding values: $(0)\rightarrow (f)$ while preserving their
entropy $\Delta S_{i}$ (due to isentropic expansion) and particle
number $\Delta N_{i}$ (because of
chemically frozen evolution). Then, according to the thermodynamic identity (%
$\mathsf{p}$ is pressure)

\begin{equation}
\Delta S_{i}=\frac{\mathsf{p}_{i(0)}\Delta V_{i(0)}^{\ast }}{T_{i(0)}%
}+\frac{\Delta E_{i(0)}^{\ast }}{T_{i(0)}}-\frac{\mu _{i(0)}}{T_{i(0)%
}}\Delta N_{i}=\frac{\mathsf{p}_{i(f)}\Delta V_{i(f)}^{\ast }}{%
T_{i(f) }}+\frac{\Delta E_{i(f)}^{\ast }}{T_{i(f) }}-\frac{\mu
_{i(f)}}{T_{i(f) }}\Delta N_{i}  \label{thermod}
\end{equation}
In the ideal Boltzmann gas $\mathsf{p}V^{\ast }=NT$ and at $m/T \gg 1$ $E\approx mN+%
\frac{3}{2}NT$. In the ultrarelativistic case $m/T\ll 1$ the
particle mass can be neglected and $E\approx 3NT.$ As the result
of  Eq. (\ref{thermod}) , in  both cases the
''non-relativistic fugacities'' will be conserved, $\frac{\mu _{i(0)}-m}{T_{i(0)%
 }}=\frac{\mu _{i(f)}-m}{T_{i(f)}}$, along the world lines of each
fluid element. What  happens to the APSD during the system
evolution? In the case of  relativistic expansion the
averaging in (\ref{APSD-tot}) should be done in relativistic covariant form $%
d^{3}rd^{3}p$\textbf{$\rightarrow $}$d\sigma _{\mu }p^{\mu }\frac{d\mathbf{%
^{3}}p}{p_{0}}$ , $f(t)\rightarrow f(\sigma ),$ $f(t,v_{L})
\rightarrow f(\sigma ,y)$ where the hypersurface $\sigma $
generalizes the hypersurface of constant time $t.$ The $4$-vector
formed by the world lines of the $i-$fluid element crossing
$\sigma $ and moving with $4$-velocity $u^{\mu }$ is $\int_{\Delta
V_{i}^{\ast }}d\sigma ^{\mu }$ $=\frac{\Delta V_{i}^{\ast
}n_{i}^{\mu }}{u_{(i)\nu }n_{(i)}^{\nu }} $, where $n^{\mu }(x)$
is normal to $\sigma (x)$. Then ($k=1,2,...$)

\begin{equation}
 \int_{\Delta
V_{i}^{\ast }}d\sigma ^{\mu } \int  p_{\mu }\frac{d\mathbf{^{3}}p}{p_{0}}%
\exp [-k(p_{\mu }u^{\mu }-m)/T]\xi ^{k}   \label{thermodyn1}
\end{equation}
 $\sim T_{i}^{3/2}\xi ^{k}\Delta V_{i}^{\ast }$ if $m/T\gg 1$ and $\sim
T_{i}^{3}\xi ^{k}\Delta
V_{i}^{\ast }$ if $m/T\ll 1$. Since the ''non-relativistic fugacity'' $%
\xi =$ exp$[\frac{\mu -m}{T}]$ and corresponding values ($%
mT_{i})^{3/2}\Delta V_{i}^{\ast }$ ( $m/T\gg 1$) or
$T_{i}^{3}\Delta
V_{i}^{\ast }$ ($m/T\ll 1)$ are constants along the current line, then

\begin{equation}
\int_{\Delta V_{i(0)}^{\ast }}d\sigma _{(0)}^{\mu } \int p_{\mu }\frac{d%
\mathbf{^{3}}p}{p_{0}}f^{k}=\int_{\Delta V_{i(f)}}d\sigma
_{f}^{\mu } \int  p_{\mu }\frac{d\mathbf{^{3}}p}{p_{0}}f^{k}
\label{thermodyn2}
\end{equation}
and, finally, the values of
phase-space densities $\left\langle f(\sigma )\right\rangle $ and $%
\left\langle f(\sigma ,y)\right\rangle $ averaged over $d\sigma
_{\mu }p^{\mu }\frac{d\mathbf{^{3}p}}{p_{0}}$ are the same on
different hypersurfaces. In  other words, the totally averaged PSD
of the relativistically expanding Boltzmann gas is approximately
preserved during the system's evolution if the mass of the
particles is much higher or smaller than the temperature of the
system.

The non-relativistic Bjorken-like model (\ref{F-def}) gives an
example of such a conservation of the APSD,   resulting in
$m\left\langle
f(t,v_{L})\right\rangle \mathbf{=}\frac{n_{0}}{8\sqrt{\pi }(mT_{0})^{3/2}}%
=const$. The pion APSD $\left\langle f(t,y)\right\rangle $ in the
relativistic chemically frozen 3D Bjorken model \cite{A-B-S} with
the initial conditions as in Table 1 is also roughly  constant: it
 changes from the chemical to kinetic freeze-out stages $5 - 7$
per cent only, but it increases with time even without feeddown
from resonance decays! This means that the non-relativistic
fugacity of pions, $\exp [\frac{\mu _{\pi }-m_{\pi }}{T}]$, grows
with time. This paradoxical behavior of the APSD is, however,
typical for a mixture of heavy and light hadron gases. Indeed, let
us imagine that we have a thermal mixture of  two gases: massless
(e.g., neutrino) with concentration $\kappa_{1}$ and massive
(e.g., neutrons) with concentration $\kappa_{2}$: $\kappa_{1} +
\kappa_{2}=1$. Each of them will preserve its non-relativistic
fugacity if they expand separately. For evolution of the mixture
we have, generalizing (\ref{thermod}), that $\kappa _{1}\mu
_{1}/T+\kappa _{2}(\mu _{2}-m)/T=const$ along the current line at
isentropic expansion. One could suppose that the expansion is
conditioned mainly by the heavy component: ($mT_{i})^{3/2}\Delta
V_{i}^{\ast }$ $\approx const$ during the evolution of the i-th
fluid element. If then both fugacities are preserved then the
entropy of the mixture will decrease: the light component needs
the volume
$\Delta V_{(f)}^{\prime }=\Delta V_{(0)}\frac{T_{(0)}^{3}}{%
T_{(f)}^{3}}$ to preserve  entropy at the same fugacity, but it
gets only $\Delta V_{(f)}=\Delta
V_{(0)}\frac{T_{(0)}^{3}}{T_{(f)}^{3}}\frac{\Delta V_{(0)}}{\Delta
V_{(f)}}<\Delta V_{(f)}^{\prime },$ while the heavy component
preserves its entropy. Thus, the fugacity of the light component
should grow and, correspondingly, the fugacity of the heavy
component should decrease. The
expansion of the mixture then follows a law which is intermediate between $%
V\sim 1/T^{3/2}$ and $V\sim 1/T^{3}$ . Such  behavior was
considered in detail in Ref. \cite{A-B-S}. The mean mass of the
chemically frozen hadron-resonance Boltzmann gas is significantly
higher than the temperature of hadronization:
$\overline{m}=\sum \kappa _{j}m_{j}=0.662$ GeV, where $%
\kappa _{j}$ is the concentrations of $j-$hadronic species. The
aforementioned result on the increase of the pion APSD
$\left\langle f(t,y)\right\rangle $ during the chemically frozen
hydrodynamic evolution \cite{A-B-S} is explained, therefore, by
the fact that the pion gas is the lightest component in the
mixture and so the pion non-relativistic fugacity should grow
regardless of inelastic rescatterings  and resonance decays.

\section{Resonance influence on spectra and interferometry volume in A+A
collisions}
\subsection{Emission and decays of resonances in quasi-classical approximation}
The above analysis is based on thermal local equilibrium
distributions which in hydrodynamic models describe so called
direct particles and resonances. The pion spectra from resonance
decays are not thermal and, even under an assumption of sharp
kinetic freeze-out, are formed in some effective post freeze-out
4-volume that depends also on resonance widths. Unfortunately, one
cannot ignore \textit{a priori} the contributions to the pion
spectra and interferometry radii by the short-lived resonances as
they give more than half of total pion yields in ultrarelativistic
A+A collisions. Such a big contribution arises since the
chemically equilibriated composition of the hadron resonance gas,
corresponding to relatively high hadronization temperature, is
nearly frozen during the subsequent evolution to low temperatures.
Of course, the resonance decays could take place during the
evolution of the dense system and lead to some increase of number
of (quasi) stable particles even before the kinetic freeze-out. On
the other hand, there are back reactions that produce the
resonances in the dense system. These reactions do not lead to
chemical equilibrium, but just reduce the effect of resonance
decays into chemically non-equilibriated media. According to the
results of Ref. \cite{A-B-S}, nearly $2/3$ of all pions arise from
the resonance decays after the thermal freeze-out if one totally
ignores the reduction of resonance numbers during the hydro
evolution. Taking into account a rather short duration of the
hydrodynamic stage for the hadron gas, about $1.7$ fm/c, which is
comparable with the average life-time of the resonances, we can
very roughly estimate that at least  half of the pions arise from
resonance decays during the post hydrodymanic stage.

The only attempt to analyse the contribution of large  numbers of
short-lived resonances to the spectra and interferometry volume
was done in Refs. \cite{Krakow} within some hydrodynamic
parametrization of the kinetic freeze-out. The aim of those papers
was to fit the spectra and interferometry data from  CERN  SPS and
BNL RHIC supposing that kinetic freeze-out happens just after
hadronization. However,  the analysis of \textit{relative}
resonance contribution to the interferometry radii has not been
done probably since  a non-physical approximation corresponding to
infinite widths (zero lifetimes) of all resonances was used. But
such estimates are necessary for an analysis of the phase-space
and particle density properties at  kinetic freeze-out. Pursuing
this aim we analyse the resonance contributions to the spectra and
interferometry radii based on the hydrodynamic model of Ref.
\cite{A-B-S} and take the widths $\Gamma $ of mesonic and baryonic
resonances with masses up to 2 Gev from Particle Data Group
\cite{Data}. Because of  technical reasons we ignore, however, the
contributions from many-cascade decays. We hope these do not
change the general picture and conclusions. As we show below, the
single particle transverse pion spectra have the same main
features as in the model accounting for the many-cascade decays of
heavy resonances \cite{Krakow}.

To evaluate the contribution from resonance decays to the pion
spectra and interferometry radii we use the quasi-classical
approximation \cite{Bolz1,Heinz1} for the emission function
$g_{i}(x,p)$ of the i-th resonance species:
\begin{eqnarray}
g_{i}(x,p) &=&\int\limits_{\omega ^{-}}^{\omega ^{+}}d\omega ^{2}\varphi
_{i}(\omega ^{2})\int \frac{d^{3}p_{i}}{E_{i}}2m_{i}\delta
((p_{i}-p)^{2}-\omega ^{2})\cdot   \nonumber \\
&&  \label{reson} \\
&&\int\limits_{0}^{\infty }d\tau ^{\ast }\Gamma _{i}\exp (-\Gamma _{i}\tau
^{\ast })\int\limits_{\sigma }d\sigma _{\mu }(x_{i})p_{i}^{\mu }\delta
^{(4)}(x^{\mu }-[x_{i}^{\mu }+\frac{p_{i}^{\mu}\tau ^{\ast }}{m_{i}}%
])f_{i}^{l.eq.}(x_{i},p_{i})  \nonumber
\end{eqnarray}
The emission function $g_{i}(x,p)$ describes the production of
pions with momentum $p$ from the i-th resonance species, the
latter being distributed
according to the local equilibrium phase-space density $%
f_{i}^{l.eq.}(x_{i},p_{i})$. The quasi-classical picture supposes
that the i-th resonance is created from the hydrodynamic tube at a space-time point $%
x_{i}^{\mu }$ of the kinetic freeze-out hypersurface $\sigma $ and decays
into $\pi +X$ after some proper time with the mean value $1/\Gamma _{i}$.
The probability to produce a group of particles $X$ with an invariant mass $%
\omega $ which accompany the pion at the decay of the i-th
resonance is described by the covariant amplitude of the resonance
decay $\varphi _{i}(\omega ^{2})$ that is supposedly  known. The
pion spectra and correlations are defined by the phase-space
densities of thermal (''direct'') pions and resonances at $\sigma
$. According to Ref. \cite{A-B-S},  the local Boltzmann
distribution $f_{i}^{l.eq.}(x_{i},p_{i})$ has a transverse
Gaussian-like density profile with width $R(\tau _{th})$, and is
attributed to the kinetic freeze-out hypersurface $\sigma $: $\tau
_{th}=const$
 where the system is characterized by the constant
temperature $T_{th},$ boost-invariant longitudinal flow and
transverse flow $y_{T}=v_{R}r/R(\tau _{th})$.  The numerical
values of correspondent parameters for Pb+Pb 158 AGeV collisions
were extracted in Ref. \cite{A-B-S} from an analysis of the
spectra and interferometry radii at fairly large $p_{T}$ to
minimize  the influence of resonance decays. They are
$T_{th}=0.135$ GeV, $R(\tau _{th})=5.3$ fm, $v_{R}=0.4.$ We take
proper freeze-out time $\tau _{th}=8.2$ fm/c for the ''direct''
pions \cite{A-B-S}, and use this freeze-out time also for
resonances which produce pions. Note, however, that the average
freeze-out time for different hadronic species was found to be
$8.9$ fm/c \cite{A-B-S} which  does not differ much from what we
use. We do not intend in this paper to perform a detailed fitting
of the spectra and radii; our aim is to analyse the dependence on
flow intensity of the relative contributions of resonances to the
spectra and interferometry radii within the appropriate model
\cite{A-B-S} and, thus, to estimate the influence of resonances on
the phase-space densities. So, we just take the parameters which
were found in \cite{A-B-S}, and use them for highest SPS energies.
Also, we estimate roughly the correspondent contribution to the
APSD at RHIC energies by using the \textit{same} model, simply
enhancing the intensity of transverse flows: $v_{R}=0.4\rightarrow
v_{R}=0.6$.
\subsection{The results on spectra and interferometry and their interpretations}
The analysis of the single pion spectra from decays of the fixed
resonance species shows it is essentially non-thermal,
non-exponential, at least in the transverse momentum region below
$p_{0}$. One can see it at the top of Fig. 4 for the fixed heavy
resonance species $f_{2}(1270)$ with $p_{0}=0.622$ GeV.
Nevertheless, the total contribution (thermal + decays) from many
 such resonances with different masses and different $p_{0}$
results in a slope of the transverse spectrum which, as one can
see from Fig. 4, is surprisingly similar to the spectrum slope for
the thermal ("direct") pions in the wide momentum region at
different intensities of transverse flows. The same peculiarities
of the resonance contributions to the transverse pion spectra were
found also in Ref. \cite{Tarasov} within the Landau hydrodynamic
model with realistic freeze-out conditions. In contrast to these
observations, the results of Refs. \cite{Krakow} demonstrate a
noticeable decrease of the effective temperature, or inverse of
the slope of total spectra, as compared to  thermal pions. The
main reason for such a discrepancy is the (unrealistically) high
decoupling temperature $T$ that is taken to be equal to the
hadronization, or chemical freeze-out one in Refs. \cite{Krakow}.
It is qualitatively clear that the ratio of effective temperatures
of total to thermal pion spectra depends on the relation between
the mean $p_0$ and $T$. We demonstrate this effect in Fig. 5 where
one can see that the slopes of the total pion spectra are about
the same as they are for thermal pions  in the realistic interval
of freeze-out temperatures $100 - 140$ MeV. Above this region the
pion effective temperatures for the total transverse spectra are
significantly smaller than for just thermal spectra. The result
corresponding to the decoupling temperature equal to the chemical
freeze-out one, $163$ MeV \cite{A-B-S}, is close to what was found
in Refs. \cite{Krakow} for similarly high temperatures
$T\approx165$ MeV. Clearly, at  freeze-out temperatures below
 $100 - 140$ MeV the transverse effective temperature of total pion
production has to be noticeably larger than in the case of thermal
pions only. The correspondent tendency one can perceive from the
plots in Fig. 5.

Let us move to a discussion on the influence of resonance decays
on Bose-Einstein correlations. Fig. 6 demonstrates that resonance
decays extend the interferometry volume, especially at small
transverse momenta. The ratio of the complete volume to the direct
one grows when the intensity of transverse flow increases. It
could be the double-effect of an increase in the resonance
velocities in the c.m. of the system and correspondent Lorentz
dilation for the resonance life-times; both effects increase the
mean free path for outgoing resonances. Another impressive effect
 of the flow and resonance decays on the interferometry
radii is demonstrated in Fig. 7. One can see that the ''standard''
hydrodynamic behavior of the ratio $R_{out}/R_{side}$ which is
typically larger than unity\footnote{Except for an (artificial)
situation with a box-like density profile \cite{Blast}.} becomes
essentially smaller, $R_{out}/R_{side}<1$, outside  the region of
small $p_{T}$ if the resonance decays are taken into account. The
above effects are created by the rather complicated interplay
between the kinematics of resonance decays and flows. A
description and understanding  of the correspondent effects are
crucial for the correct estimates of the phase-phase densities.

Let us consider the effect of the relativistic longitudinal flow.
If there is no flow and the resonance momentum spectra are, for
instance, pure thermal, then all the  points of the whole source
volume are  accessible  with equal probability for  resonance
emission. Since the resonances have  finite average life-times,
they can decay inside as well as outside of the source volume,
thereby, at any momenta  the pion interferometry radii will be
larger than the size of the source that emits the resonances. If
the source expands relativistically in the longitudinal direction,
then the spectra of \textit{heavy} resonances are defined mainly
by the flow, not the temperature. Let us suppose that pairs of
identical pions with average rapidity $y\simeq 0$ and half-sum of
pion transverse momenta $ p_{T}$  are produced by two resonances
$1^{\ast }$ and $2^{\ast }$ having, for simplicity, only two-
particle modes of decay (pion plus, e.g., nucleon) with the same
absolute value of pion momenta $p_{0}$ in the rest frame of each
resonance. If one studies the \textit{long-}projection of the
correlation function, the transverse momenta of both pions is
supposed to equal $\mathbf{p}_{T}$ . Since  only pions with close
momenta interfere, the resonances should have rapidities
$y_{1}^{\ast
}\simeq \pm y_{2}^{\ast }$ $=\pm \sinh ^{-1}\sqrt{\frac{p_{0}^{2}-p_{T}^{2}}{%
p_{T}^{2}+m_{\pi }^{2}}}$  to guarantee $y\simeq 0$, if the
resonance masses are much bigger than the temperature and there
are no transverse flows. Thus, the heavy resonances with
$y_{1}^{\ast }=\pm y_{2}^{\ast }$ are emitted just from the two
fluid elements with correspondent fluid rapidities $\eta
_{i}=y_{i}^{\ast }$, see Fig. 8. In the case $y_{1}^{\ast }\simeq
y_{2}^{\ast }$ the fluid elements are overlapping. The
sizes $l$ of these fluid elements are rather small, $l\simeq\tau \sqrt{\frac{T}{%
m_{i}^{\ast }}}$ \cite{Si}, as compared to the system's scales and
also to the correspondent homogeneity lengths of direct pions. If
the half-sum of pion transverse momenta is small, $p_{T}\ll
p_{0}$, then the distance between the two fluid elements could be
also small if $y_{1}^{\ast }=y_{2}^{\ast }$ or fairly large, $L\gg
l$, if $y_{1}^{\ast }=-y_{2}^{\ast }$ . The former configuration
corresponds to situations a) and b) in Fig. 8 at the moment of
resonance decays $\tau >\tau_{f.o.}$. The later configuration c)
increases significantly at  small $p_{T}$  the \textit{long-
}interferometry radius, that corresponds to a Gaussian fit to the
correlation function. It is enhanced additionally since in this
case the velocities of the resonances are back-to-back directed
(total momentum of pair is selected to be  near zero), and they
propagate for some time until they decay. At large $p_{T}>p_{0}$
the distance between the two resonances is reduced according to
the kinematics, especially when transverse flow takes place, which
is why the resonance contribution to the interferometry volume
dies out with $p_{T}$ much faster than their contribution to the
spectra, see Fig. 6.

Let us illustrate also the reduction of the ratio
$R_{out}/R_{side}$ due to heavy resonance  decays in the presence
of transverse flows. For  simplicity we omit the longitudinal
coordinate and will keep in mind the linear with  transverse
radial flow and the Gaussian-like particle density profile at
freeze-out. Let us consider at $p_T=p_0$ the interferometry in
transverse plane for the pions coming, as we described above, from
the one sort of resonances characterized by $m_i^{*}\gg T$ and
with pion decay momentum $p_0$. In Fig. 9 we mark the four points
a, b, c, d of emission of the resonances at freeze-out which are
minimal and maximal in $x$ (out-) and $y$ (side-) directions at
the given restriction for pion momenta. Those positions are
dictated by the kinematics of the decay ($v_0 = p_0/E_{\pi}$) to
guarantee the pion transverse momentum to be $p_T=p_0$ and
directed along the $x$ axis. As one can see from Fig. 9, the
maximal sideways extension $R_{bd}$ (along the $y$ axis) of the
emission region for the resonances producing the pions  is equal
to the  maximal outward extension $R_{ac}$ of the region, and the
maximal relative resonance velocities in the $x$-direction $\Delta
v_{x}=2v_0$ is equal to the maximal relative velocities in the
$y$-direction $\Delta v_{y}=2v_0$. If the density were constant,
the effective emission sizes would be  $R_x = R_y$. But since the
density decreases with the radius, the effective outward size of
the emission region, accounting for the probabilities of emission,
will be less, $R_x<R_y$, since point $R_c$ is maximally remote
from the centre as one can see from Fig. 9. Then we have also
$\Delta v_{x} < \Delta v_{y}$ effectively. It reduces the ratio of
the interferometry out- to side- radii especially when mean
life-time $\tau^{*}$ of the resonances are fairly large: $R_{out}
=R_x + \tau^{*}\Delta v_{x} < R_{side}=R_y + \tau^{*}\Delta
v_{y}$. The results of Fig. 7 demonstrate the total effect from
the contributions of many resonance species in the 3D expanding
hydrodynamic model.

Note that the above effects are typical in the  hydrodynamic
picture with sudden freeze-out when the fraction of pions produced
by resonances is fairly big as  takes place in the case the
chemical freeze-out. In the opposite case, mainly direct-direct
and direct-resonance pion pairs are important for interferometry
radii formation; the correspondent results are well known
\cite{Bolz1,Heinz1}.
\subsection{Resonance decays and phase-space densities}
This detailed consideration allows us to clarify what we should
measure to estimate the overpopulation of phase-space in the
system at the latest stage of its evolution. First, the spectra of
pions from decays of some fixed resonance species are primarily
non-thermal; second, the resonances that produce the pion pair are
emitted, in general, from \textit{different} small  elements of
fluid separated by some distance that could be rather large at
small transverse momenta of the pair. The pions of the pair are
produced also at different (proper) times. This is completely
unlike the emission of direct pions at  freeze-out in the
hydrodynamic picture: the pions with similar momenta $p$ are
emitted  from the same fluid element, the value of momenta $p$
dictates the size of that element (the homogeneity length) and the
HBT interferometry measures the correspondent size. Hence, the
maximal phase-space density at each $p$ is just the spectra $n(p)$
divided by the interferometry volume and the APSD is less than it
by a factor $(2)^{3/2}$ if one supposes a Gaussian phase-space
density profile (at fixed $p$ !) outside of that fluid element.
Such a quantity, the APSD,  for the case when $x-p$ correlations
are absent and the Gaussian profile is universal for all $p$ was
analysed in the pioneer Bertsch paper \cite{Bertsch}. For pions
created by the resonances there is no such  transparent physical
meaning of the ratio of the spectrum to interferometry volume,
even if  the resonance width is set to infinity and so they are
only allowed to produce pions  at the freeze-out hypersurface.
Indeed, if we suppose the absence of transverse flows,  then at
small $p_{T}$ of pairs there are, as was discussed above in detail
(see also Fig. 8), the two rather small, $l\sim \tau
\sqrt{T/m^{\ast }}$, and well separated by distance $L\sim 2\tau
\sinh \eta _{1}\gg l$, fluid elements which emit the pions by
means of resonances. Obviously, the maximal APSD here is
proportional to $%
n(p)/l$ and is a high value, while it will be estimated as a small value, $%
\sim n(p)/L$, according to the Bertsch prescription, when the
interferometry radius,  extracted by a Gaussian fit to the
interferometry peak, is proportional to $L$ . Of course, the
correspondent correlation function in a large interval of $q_{L}$
is rather different than the Gaussian form; it is proportional to
$1+\exp (-q^{2}l^{2})\cos {}^{2}(qL/2)$ , and this is the solution
of the paradox - the simplest form, the Bertsch prescription, just
fails here, but how does one distinguish such  effects of the
non-Gaussian contribution, when experimenters are fitting  the
\textit{total} correlation function from many  similar
contributions to a Gaussian? The finite life-time of the
resonances makes the situation even more complicated. We can
conclude that the effective region (homogeneity volume), where the
single pion spectrum from the resonance decays forms, is not
associated with an interferometry volume of pion ''resonance
emission'' that is topologically non-trivial and corresponds, in
general, to the superposition of space-time separated Gaussian
sources. In the next Section  we propose a way  to extract the
actual APSD of pions in such a situation.

\section{ Extraction of pion fugacities in
A+A  collisions.}
\subsection{Sudden freeze-out vs continuous emission: how to treat the APSD?}
 Our aim is to propose a simple method to estimate the
overpopulation of phase-space that does not depend on the details
of hydrodynamic evolution. Before discussing it, let us make some
comments on the applicability of the hydrodynamic approach to the
post hadronization stage in A+A collisions. As we discussed in
Sec. 2, there is approximate equivalence between the chemically
frozen hydrodynamics of a hadron gas and its evolution within the
cascade approach in the temperature region of applicability of
hydrodynamics. Out of this region, below the temperature 0.12 GeV
\cite{Teaney}, the formation of spectra is basically continuous in
time with fairly long
''tails''. That continuous emission is characterized by the emission function $%
g(t,\mathbf{r,p})$ which, normally, is very far from thermal. The
 example of it is presented in Ref. \cite{Sinyukov} and is
based on the exact analytic solution (\ref{hydro})  of the
Boltzmann equation for an expanding fireball. Unlike the very
complicated structure of the emission function $g(t,\mathbf{r,p})$
that was found in \cite{Sinyukov}, the interferometry radii,
spectra and the APSD have simple and clear analytical forms and
correspond to the thermal distribution function in the fireball
before it starts to decay. It looks like there is some kind of
duality \cite{duality} in the description of the spectra and the
interferometry data based either on the (thermal) distribution
functions $f(t,\mathbf{r,p})$ which characterize the system just
before decay begins or on the emission function
$g(t,\mathbf{r,p})$ that describes the process of continuous
particle emission after the system loses its simple thermal
properties. The replacement  of the complicated emission process
by the simple Landau \cite{Landau} prescription of sudden
freeze-out at the lower boundary of the region of applicability of
hydrodynamics, at $T\approx 0.1 - 0.14$ GeV (depending on the
system size) is, of course, a rather rough approximation.
Nevertheless, as we discussed above, the momentum-energy
conservation laws, particle and entropy (partial) conservation as
well as some symmetry properties minimize the correspondent
deviation for observables. As for the average phase-space
densities, which, as we discussed in Sec. 2, are approximately
frozen during the hydrodynamic evolution, there is no reason to
expect that they are changed significantly during  the transition
to their complete freeze-out (the APSD does not change during the
stage of free streaming).

Therefore, one can consider the analysis of the APSD in two
aspects. The first one associates the observed APSD with the
corresponding value \textit{ in the thermal hadron medium} just
before its decay, that allows one to obtain, based on the
distribution function $f(t,\mathbf{r,p})$, the particle number and
other densities in the medium at the pre-decay stage. On the other
hand one can link the APSD with the properties of the
\textit{particle continuous emission}, and thus  analyse the
process of particle liberation described by the emission function
$g(t,\mathbf{r,p}).$ If the decay of the medium starts at  time
$t_{f}$, then one can compare the particle density just before
decay $ n(t_{f},\mathbf{r)}=\int
d^{3}pf(t_{f},\mathbf{r},\mathbf{p})$ with the density of
''emission capacity'' of the source at point $\mathbf{r},$
$n_{emit}(\mathbf{r})=\int_{t_{f}}^{\infty }dt\int d^{3}p$
$g(t,\mathbf{r,p})$. Since the system continues to expand during
the process of particle liberation, then $n(t_{f},\mathbf{r)}\geq
n_{emit}\mathbf{(r)}$ and the equality is reached only in the case
of a \textit{real } sharp freeze-out
when, in the non-covariant form, $g(t,\mathbf{r,p})=f(t,\mathbf{r},%
\mathbf{p})\delta (t-t_{f}).$ Thus the  results of our analysis of
the APSD based on the hydrodynamic picture with the Landau
freeze-out prescription gives the possibility, on the one hand, to
find the particle density \textit{in the thermal hadron medium}
just before its decay and, on the other hand, to make the upper
estimate of the  ''emission capacity'' density for the continuous
process of particle liberation.

It is clear, however, that the resonance decays at the post
freeze-out stage destroy the above conception of duality in the
APSD description since they seriously affect  the
post-hydrodynamic (pion) emission function $g(t,\mathbf{r,p}).$
The attempt to attribute the decays of resonances just to the
sharp freeze-out hypersurface will make an estimate of the pion
density to be artificially high. Indeed, the proper life-time
$1/\Gamma$ of resonances averaged over all the species is nearly
$2$ fm/c that gives us, taking into account the Lorentz-dilation
of life-times of resonances in  the expanding swarm, an even
larger value, say 2.5 fm/c, when the total number of resonances
drops by a factor of $1/e$. Since at the last quasi-inertial
3D-stage of expansion the system volume grows roughly proportional
to $\tau^3$, one can easily estimate, keeping in mind the typical
(proper) time 10 fm/c for the freeze-out in A+A collisions at SPS
and RHIC, that despite the resonance decays bringing about  half
of the pion yields the pion densities nevertheless decrease  at
the post freeze-out stage! It means that at the final decoupled
stage of the system evolution the maximal densities, including the
APSD of pions, are reached  at the thermal freeze-out, at  the
boundary of the applicability region of the hydrodynamic approach.
Thereby, taking into account also that resonance contributions to
the APSD is physically obscure and their including complicates the
treatment of the APSD very much (see the discussion at the end of
Sec. 3) we will exclude the resonance contributions to the spectra
and interferometry radii when estimating the maximal pion APSD.
 Since the effective temperatures of the total pion spectra and
direct pions are approximately the same until rather high $p_{T}$
, one can just reduce the spectra by the corresponding factor,
$1/3 - 1/2$, in accordance with the chemical freeze-out concept
\cite{Braun-Munzinger,A-B-S}. As for the interferometry volume,
the correction factor can be taken as $0.6$ for SPS energies and
$0.5$ for RHIC energies as  follows from Fig. $4$ and will be
explained below.
\subsection{A model independent method extracting fugacity}
 Coming back to a hydrodynamic analysis of the spectra, interferometry
radii and the APSD of thermal particles in Sec. 2, one can see
(e.g. from Eqs. (\ref{spect-sol2}), (\ref{PSDA}) ) the simple link
between the above values in the non-relativistic hydrodynamic
models
\begin{equation}  \label{simple}
\left\langle f(t,p)\right\rangle =\frac{d^{3}N}{d^{3}p}/(8\pi ^{3/2}V_{int})
\end{equation}
Eq. (\ref{simple}) expresses the basic Bertsch idea \cite{Bertsch}
and can be generalized in covariant way for a relativistic situation \cite{Heinz2}%
. If a system decays on some freeze-out hypersurface $\sigma $,
the PSD averaged over this hypersurface can be expressed through
the two-particle correlation function $C(p,q)$ of identical bosons
as the following
\begin{equation}  \label{covariant}
\left\langle f(\sigma ,p)\right\rangle =\frac{\int \left( f^{l.eq.}(t,%
\mathbf{r},p)\right) ^{2}p^{\mu }d\sigma _{\mu }}{\int f^{l.eq.}(t,\mathbf{r}%
,p)p^{\mu }d\sigma _{\mu }}=\frac{1}{(2\pi )^{3}}\int \frac{d^{3}N}{d^{3}p}%
(C(p,q)-1)d^{3}q
\end{equation}

Thus, to estimate the APSD one should use the experimental data on
spectra and correlation functions or just the interferometry radii
if $C(p,q)-1$ is represented in standard Gaussian form, see
(\ref{simple}). As for the left hand side of Eq.
(\ref{covariant}), it is strongly model dependent since it is
determined by unknown functions in space-time such as the fugacity
$\xi =\exp (\beta \mu )$, temperature $T\equiv 1/\beta $ and flow
4-velocities $u^{\mu }$ in the local equilibrium Bose-Einstein
distribution $f(x,p)=(2\pi )^{-3}[\xi ^{-1}\exp (\beta p\cdot
u)-1]$.  More concretely, it was demonstrated in Ref.
\cite{Tomasik} that boost-invariant longitudinal flows in the
system could bring  significant error if the fugacity is extracted
from $\left\langle f(p)\right\rangle $ assuming a static source
\footnote{ Note,  it is erroneously claimed in Ref. \cite{Tomasik}
that such a  discrepancy vanishes at high $p_{T}$, while the
fugacity is still a factor of $\sqrt{2 }$ larger (in the Boltzmann
approximation) at high $p_{T }$ than one could  estimate supposing
that the source is static.}. It was shown also that the extracted
value of fugacity depends strongly on transverse flows - all of
which  makes estimates of this value  essentially model dependent.
Therefore the experimental analysis (see, e.g. \cite{Ray}) of the
overpopulation of phase space is carried out within  some concrete
model and, thus, depends on many assumptions  such as distribution
functions on $\sigma $ and the form of this freeze-out
hypersurface in Minkowski space.

Here we propose a method that allows one to extract the fugacity
in a model independent way. Let us explain the basic idea using
the oversimplified example when the  freeze-out happens at
uniform temperature and uniform particle densities across the
\textit{whole} freeze-out hypersurface enclosing  the space-time
of the expanding hydrodynamic system. Then the phase- space
density averaged over the hypersurface and momentum,
\begin{equation}
\left\langle f\right\rangle =\frac{\int \frac{d^{3}p}{p_{0}}p^{\mu }d\sigma
_{\mu }f^{2}(x,p)}{\int \frac{d^{3}p}{p_{0}}p^{\mu }d\sigma _{\mu }f(x,p)}=%
\frac{\int \left( \frac{d^{3}N}{d^{3}p}\right) ^{2}(C(p,q)-1)d^{3}pd^{3}q}{%
(2\pi )^{3}N},  \label{complite}
\end{equation}
will be a suitable value which allows one to extract the fugacity
at the  freeze-out unambiguously, no matter to the form of the
hypersurface and flow profile. Indeed, at each point $x$ one can
make the integration over $\mathbf{p}$ in the rest frame of the
fluid element at the point $x$ taking into account that in this
comoving system the $f(x,p)$ depends on the absolute value of
$\mathbf{p}$ only (property of local equilibrium). Then, coming
back to  the common c.m. system, and exploring the
Lorentz-invariant properties of the local equilibrium distribution
function, one gets
\begin{equation}
(2\pi )^{3}\left\langle f(\sigma )\right\rangle =\frac{\int d\sigma _{\mu
}u^{\mu }(r,\eta )}{\int d\sigma _{\mu }u^{\mu }(r,\eta )}\frac{\int \frac{%
d^{3}p}{(\xi ^{-1}(r,\eta )\exp (\beta (r,\eta )p_{0})-1)^{2}}}{\int \frac{%
d^{3}p}{\xi ^{-1}(r,\eta )\exp (\beta (r,\eta )p_{0})-1}}  \label{dynamic}
\end{equation}
If, as we supposed, the temperature and particle densities are constant at
freeze-out, then
\begin{equation}
(2\pi )^{3}\left\langle f(\sigma )\right\rangle =\frac{\int \frac{d^{3}p}{%
(\xi ^{-1}\exp (\beta p_{0})-1)^{2}}}{\int \frac{d^{3}p}{\xi ^{-1}\exp
(\beta p_{0})-1}}  \label{static}
\end{equation}
and $\left\langle f(\sigma )\right\rangle =$ $\left\langle
f\right\rangle _{eq}$ where $\left\langle f\right\rangle _{eq}$ is
just the APSD for emission of a static thermal homogeneous source.

Despite the extreme simplicity of this result, it cannot be used
practically  for two reasons. The first  is that the
integration in Eq. (\ref{dynamic}) must be done over the whole \textit{closed }%
freeze-out hypersurface, including, of course, butt-ends of the
hydrodynamic tube. It implies the knowledge of one- and
two-particle spectra in the whole momentum region ($y,p_{T}$)
which  cannot be obtained by modern detectors. The second
important reason why the formula $\left\langle f(\sigma
)\right\rangle =$ $\left\langle f\right\rangle _{eq}$ cannot be
used is that the baryochemical potential and, hence, the
temperature are not uniform across the longitudinal direction as
the data on proton to antiproton ratios in different rapidity
regions demonstrates \cite{Exp}; see also the discussion about
inhomogeneities  of thermodynamic parameters in rapidity in Refs.
\cite{A-B-S}, \cite{Sollfrank}.

Fortunately, a result similar to (\ref{dynamic}) can be obtained
for the central rapidity interval where the spectra are usually
measured and the thermodynamic parameters are nearly uniform. Let
us start from the well known observation \cite{Landau} that at the
latest stage of a hydrodynamic evolution the longitudinal velocity
distribution corresponds to the asymptotic quasi-inertial regime,
$v_{L}=r_{z}/t$. It takes place even in the Landau model of
complete stopping, and, of course, in the Bjorken model
\cite{Bjorken} where this quasi-inertial regime exists at all
times of hydrodynamic evolution. Further, if deviations of
thermodynamic values such as  temperature, baryochemical
potential, etc. in the longitudinal direction near central fluid
rapidity $\eta =0$ are negligible within the
maximal rapidity length of homogeneity \cite{Si}, $\sqrt{\frac{T}{m_{\pi }}}%
\simeq $ $1$, then all physical values at mid-rapidity can be
calculated supposing  boost-invariance. The latter implies that
the distribution function which is applied   for evaluation of the
spectra at mid-rapidity near $y=0 $ can be considered to depend
only on the difference between fluid and particle rapidities,
$\eta -y$. As a result, the APSD at fixed particle rapidity
$\left\langle f(\sigma ,y)\right\rangle $ is equal to the APSD
$\left\langle f(\sigma ,\eta )\right\rangle $,
\begin{equation}
\left\langle f(\sigma ,\eta )\right\rangle \equiv \frac{\int f^{2}(t,%
\mathbf{r},p)p^{\mu }\frac{d\sigma _{\mu }}{d\eta }\frac{d^{3}p}{p_{0}}}{%
\int f(t,\mathbf{r},p)p^{\mu }\frac{d\sigma _{\mu }}{d\eta }\frac{d^{3}p%
}{p_{0}}},  \label{PSDA_y}
\end{equation}
at the fixed ratio $r_{z}/t=\tanh \eta $, corresponding to the fluid rapidity $%
\eta =y$. Then using the Lorentz transformations as described
above one can absorb all dynamic and kinematic characteristics
into common factor $ V_{eff}(\eta )=\int \frac{d\sigma _{\mu
}}{d\eta }u^{\mu }(r,\eta )$ for numerator and denominator in Eq.
(\ref{PSDA_y}). Note, if the freeze-out hypersurface $\tau
(\mathbf{r)}$ has non-space-like sectors, the measure of
integration in (\ref{PSDA_y}) should be modified to exclude
negative contributions to particle numbers at some momenta. As a
result, it contains $\theta -$functions like $\theta (p_{\mu
}n^{\mu }(x))$ where $ n^{\mu }$ is a 4-vector orthogonal to the
hypersurface $\sigma $ \cite{Sin4,Bugaev}.
 If a fluid element that crosses the time-like sector of the
freeze-out hypersurface decays preserving its total particle
number, then the measure of integration is modified according to
the prescription in Ref. \cite{Sin4}. In that case, after
integration over all momenta at each point $x$ the factorization
of the numerator  and denominator in (\ref{PSDA_y}) takes place
again resulting finally in the common  factor $V_{eff}(\eta )$ which cancels.
 Thereby, the expression (\ref{PSDA_y}) for the APSD can be used in the same form even when
freeze-out hypersurface contains non-space-like sectors, if one
considers the prescription from Ref. \cite{Sin4} to describe the
decay of the  system  on  the  time-like parts of the
hypersurface. Finally, the common effective volumes are canceled
and the result is
\begin{equation}
(2\pi )^{3}\left\langle f(\sigma ,y)\right\rangle _{y=0}=\frac{\int \frac{d%
\mathbf{p}}{(\xi ^{-1}\exp (\beta p_{0})-1)^{2}}}{\int \frac{d\mathbf{p}}{%
\xi ^{-1}\exp (\beta p_{0})-1}}  \label{result1}
\end{equation}
Here the inverse of temperature $\beta $ and fugacity $\xi $
characterize thermal properties of the decaying system \textit{at
mid-rapidity }and are supposed to be approximately constant within
one unit of rapidity near $\eta =0.$ On the other hand, the above
value of the APSD is expressed through the transverse spectra and
the interferometry radii at mid-rapidity similar to Eq.
(\ref{complite}),
\begin{equation}
(2\pi )^{3}\left\langle f(\sigma ,y)\right\rangle \approx \kappa
\frac{2\pi ^{5/2}\int \left( \frac{1}{R_{O}R_{S}R_{L}}\left(
\frac{d^{2}N}{2\pi m_{T}dm_{T}dy}\right) ^{2}\right)
dm_{T}}{dN/dy}\label{PSDA_exp}
\end{equation}
 Here we neglect  interferometry cross-terms since they are usually rather
small in the mid-rapidity region. The factor $\kappa $, which will
be calculated later from the results of the preceding Section, is
introduced to eliminate contribution of  short-lived resonances to
the spectra and interferometry radii. It absorbs also the effect
of the suppression of the correlation function due to long-lived
resonances.

Thereby we proved, using very natural assumptions, the possibility
of extracting in a model independent  way the fugacity in
expanding thermal systems at the stage of their  freeze-out,
realizing, thus, the program which was declared by Bertsch in his
pioneer paper \cite{Bertsch}.
\subsection{Pion APSDs and chemical potentials at SPS and RHIC}
Let us use our approach to find the chemical potential of direct
pions at  thermal freeze-out at  SPS and RHIC. First we have to
obtain the value of $\kappa $ in Eq. (\ref{PSDA_exp}). This
coefficient includes, in particular, the factor $\sqrt{\Lambda}$
(see, e.g., review \cite{lambda})
 arising because decays of
the long-lived ($l$) resonances such as $\eta $-, $\eta ^{\prime
}$ --mesons give the contribution $d^{3}N_{i}^{(l)}/d^{3}{\bf p}$
to the total pion spectra but not to the observed interferometry
radii. The reason is that the (theoretical) correlation function
due to pions produced by the long-lived resonances  contains very
narrow (much smaller than momentum resolution $q_{\min }$ of a
detector) peak at $q=0$ on the top of smooth curve associated with
bulk pion source. This narrow peak gives, in fact, no contribution
to integral over $q$ of $(C(p,q)-1)$ (see Eq. (\ref{complite})) in
Bertsch's method, and, therefore, results just in reduction of the
total value of the APSD. In accordance with the estimates made in
Refs. \cite{Bolz1,Soff} we choose the effective suppression factor
$\Lambda$ which measures the fraction of pion pairs containing no
pions from long--lived sources,
\begin{equation}
\Lambda(p)=\left( 1-\frac{d^{3}N^{(l)}/d^{3}{\bf p}}{%
d^{3}N/d^{3}{\bf p}}\right)^2 <1,  \label{14-def}
\end{equation}
 to be equal to  $0.8$; thereby $\kappa $ contains a factor
$\sqrt{\Lambda }\simeq 0.9$. Note, that experimental value of the
suppression factor is, normally, smaller than the above
theoretical value since the many effects, except  the decays of
long-lived particles, could suppress the measured correlation
functions. The most important among them are the single- and
two--track resolution and particle misidentification. These
effects are irrelevant to the physics of A+A collisions and we
use, therefore, theoretical value of the suppression factor
associated with long-lived resonances.

Since the slopes of the complete pion spectra and thermal pions
are very similar in a wide $m_{T}$ region (see Fig. 4), we just
reduce the experimental spectra by the ratio of the total pion
number to the number of direct pions that follows from an analysis
of the particle number ratios within the concept of chemical
freeze-out. We use the reduction factor of $3$ as  is found in
Ref. \cite{A-B-S}, also we estimate for comparison the fugacity
supposing the correspondent reduction factor to be $2$. The ratio
of the complete to ''direct'' interferometry volumes can be fitted
by a
 $const/\sqrt{m_{T}}$  function in the region shown in Fig. 4. We take into
account this momentum dependence to define the effective
correction factor to observed interferometry volume which we will
accumulate again in  $\kappa $ in Eq. (\ref{PSDA_exp}). The
correspondent contribution is $1.65$ for central Pb+Pb 158 AGeV
collisions at  CERN  SPS and  is $1.92$ for Au+Au
$\sqrt{s_{NN}}=130$ GeV at BNL  RHIC.
 Thus, $\kappa =0.9\cdot (1/3)\cdot (1.65 - 1.92)=0.5$ (SPS) - $ 0.6$
(RHIC). If one supposes the ratio of the total pion number to the
direct one to be $2$, it changes the values of $\kappa $ to
$\kappa =$ $0.65$ (SPS) - $0.7$ (RHIC). We use also the latter
values to estimate the sensitivity of the results to this  ratio.

To evaluate the APSD of direct pions (\ref{PSDA_exp}) we utilize
the following parametrization of the $\pi ^{-}$ transverse spectra
and interferometry radii.

For SPS Pb+Pb(Au) $158$ AGeV:

 The transverse spectrum is
$\frac{d^{2}N}{2\pi m_{T}dm_{T}dy}=A\exp (-m_{T}/T_{eff})$,
$T_{eff}\approx 0.180$ GeV; the midrapidity density is $\frac{dN_{\pi^{-}}%
}{dy}\approx 175$ (the NA49 Collaboration, \cite{NA49}). The interferometry radii are
 $R_{L}=C_{L}/\sqrt{m_{T}}$, $R_{S}=C_{S,1}/\sqrt{1+C_{S,2}m_{T}}$ and correspondent
numerical parameters are taken from Ref. \cite{CERES} (the CERES
Collaboration). We use the approximation $R_{O}=R_{S}$  until the
minimal measured  $p_{T}$ momentum, $p_{T}=0.125$ GeV,   and our
analytical approximation of the CERES outward interferometry radii
data for $p_{T}>0.125$ GeV.

For RHIC Au+Au $\sqrt{s_{NN}}=130$ GeV:

The transverse spectrum is $\frac{d^{2}N}{2\pi m_{T}dm_{T}dy}=A$
$(\exp (-m_{T}/T_{eff})-1)^{-1}$, $T_{eff}\approx 0.218$ GeV, and
$\frac{dN_{\pi^{-}} }{dy}\approx 249$ (the STAR Collaboration,
\cite{STAR-N}). We use here the Bose-Einstein parameterization of
the transverse spectra from Ref. \cite{STAR-N} since the
integrated rapidity density of negative pions with this fitting
function is  closer to  value $\frac{dN_{\pi^{-}} }{dy}\approx
270$ presented recently by the PHENIX Collaboration at midrapidity
\cite{phenix-new} than what one can get from the exponential
parameterization of the transverse spectra, $\frac{dN_{\pi^{-}}
}{dy}\approx 229$ \cite{STAR-N}.
 The phenomenological parameterization of
interferometry radii, $R_{L}=C_{L}/ \sqrt{m_{T}}$,
$R_{S}=C_{S,1}/\sqrt{1+C_{S,2}+C_{S,3}m_{T}}$,  and the
correspondent numerical parameters are taken from Ref. \cite{STAR}
of the PHENIX Collaboration. As one can see from Fig. $3$ of Ref.
\cite{STAR}, there is some discrepancy  in  the data on $R_{S}$
radii between the STAR \cite{st-pion} and PHENIX Collaborations
\cite{STAR}. To optimize uncertainties in the forthcoming
estimates , we  utilize for the  parameter  $C_{S,1}$ the value
$8.75$ fm that is the average between "PHENIX data motivated"
value, $8.1$ fm, and "STAR data motivated" value, $9.4$ fm,
\cite{STAR}. We use the approximation $R_{O}=R_{S}$.

The results of our calculations of the APSD according to
(\ref{PSDA_exp}) are collected in Table 2. The values found are
used then to extract the pion fugacities based on our result
(\ref{result1}). In Table $3$ we present the pion chemical
potentials and densities at different typical freeze-out
temperatures.

\begin{center}
TABLE 2. The average phase-space densities (APSD) of  negative
pions for all particles ("raw") and for thermal ones.
\bigskip

\begin{tabular}{|c|c|c|c|}
  \hline
  & $(2\pi )^{3}\left\langle f(\sigma ,y)\right\rangle$ & $\kappa$ & $(2\pi )^{3}\left\langle f(\sigma ,y)\right\rangle$  \\
  & without     & corrections     & for thermal \\
  & corrections & for resonances  & (direct) pions  \\
  \hline
 SPS      &$0.210$  & $0.5 - 0.65$ & $0.105 - 0.137$ \\
 $\sqrt{s_{NN}}=17.3$ GeV &         &                &  \\
  \hline
RHIC  & $0.240$ & $0.6 - 0.7$ & $0.144 - 0.168$\\
 $ \sqrt{s_{NN}}=130$ GeV &  & & \\
  \hline
\end{tabular}
\end{center}

\bigskip

\begin{center}
TABLE 3. The chemical potentials, "raw" and for thermal pions, and
thermal densities of negative pions vs critical densities
extracted at two typical temperatures  of the kinetic freeze-out.
\bigskip

\begin{tabular}{|c|c|c|c|c|}
  \hline
  & $T_{th}$ MeV & $\widetilde{\mu }_{\pi ^{-}}$ MeV & $ \mu_{\pi
  ^{-}}$ MeV & $ n_{\pi^{-}}$ fm$^{-3}$
\\
  & freeze-out    &  without    & for thermal &   thermal \\
  & temperature   & corrections          & (direct) pions & pion density  \\
  \hline
 SPS      &$100$  & $78$ & $32 - 51$ & $0.014 - 0.017$  \\
 $\sqrt{s_{NN}}=17.3$ GeV &         &                &  & $[0.065]$ \\
  \hline
   SPS      &$120$  & $70$ & $15 - 38$ & $0.022 - 0.027$  \\
 $\sqrt{s_{NN}}=17.3$ GeV &         &                &  & $[0.095]$ \\
  \hline
RHIC  & $100$ & $85$ & $54 - 64$ & $0.018 - 0.020$ \\
 $ \sqrt{s_{NN}}=130$ GeV &  & & & $[0.065]$ \\
  \hline
RHIC  & $120$ & $78$ & $42 - 54$ & $0.028 - 0.031$ \\
 $ \sqrt{s_{NN}}=130$ GeV &  & & & $[0.095]$ \\
  \hline
\end{tabular}
\end{center}

All the values are related to the negative pions. The values in
square brackets  in the last column of Table 3 are the critical
pion densities (in the rest frames of the fluid elements) at the
correspondent temperatures for one (negative) component of the
isospin pion triplet. As we can conclude, no exotic phenomena
associated with the overpopulation of phase-space could be
expected at the current energies. Some increase of the APSD takes
place at  RHIC; the pion chemical potentials are higher than at
SPS and correspondingly the pion number density at  RHIC is closer
to the critical value for Bose-Einstein condensation. However the
density values are still too small to expect the critical
phenomena, such as serious spreading of multiplicity
distributions, significant reduction of the interferometry radii,
etc., which are associated with the Bose-Einstein condensation of
pions in momentum space. Note, as  was discussed above, our
estimates of the pion densities at the end of the hydrodynamic
stage of the evolution are, at the same time, the upper limits for
the ''emission capacity'' density of the source of the direct
pions.

The APSDs per  pion component in the chemically equilibrated
neutral pion gas are $0.07$ and $0.09$ for $T_{th}=0.100$ GeV and
$T_{th}=0.120$ GeV respectively. As one can see from the Table 2,
the estimated APSD for thermal pions are  somewhat higher than in
the case of chemical equilibrium. Thereby some corrections to
observables conditioned by the partial overpopulation of the pion
phase-space should be taken into account in advanced models of A+A
collisions.

\section{Conclusions}

An approach which allows one to estimate the chemical potential
and, thus, the overpopulation of particle phase-space at  thermal
freeze-out in a way that does not depend on the flow profile and
form of the freeze-out hypersurface has been developed. It
realizes   Bertsch's basic idea to find from the interferometry
data the phase-space density averaged over both configuration and
momentum spaces and compare it with the correspondent value for
static source with equilibrium (homogeneous) Bose-Einstein thermal
distribution to estimate thereby the overpopulation of the
phase-space. We have proved that under the assumption of
uniformity of particle density at the freeze-out hypersurface the
flows and form of the hypersurface are absorbed into a factor that
is canceled when the APSD is determined. It is worth to note that
such a "cancellation" takes place not only for the APSD. For
example, as one can find  easily under the same conditions, the
thermal pion entropy per particle also does not depend on the form
of the hypersurface and flows.

Another problem that has been analysed is the resonance
contributions to pion phase space densities. Because of  chemical
freeze-out an essential fraction of pions, about  half, will be
produced by short-lived resonances after thermal freeze-out. It is
found that the decays of many heavy resonances correct
significantly the interferometry radii of direct pions. It could
result in up to a $50$ per cent increase of the  correspondent
interferometry volume at small transverse momenta. Since the heavy
resonances have, typically, large momentum $p_{0}$ of  pions
produced, the resonances contribute to the interferometry radii up
to  $p_{T}\sim 1$ GeV. One of the important consequences of this
is the ratio of the outward interferometry radius to the sideward
one, $R_{out}/R_{side}$, that becomes \textit{less than unity} due
to heavy resonance decays in contrast to standard results
$R_{out}/R_{side}>1$ for direct pions in the hydrodynamic models
(except for  the so-called ''blast-wave'' model \cite{Blast}, see
Footnote 1). It could be a step towards understanding  the HBT
puzzle.

The correlation function of two identical pions arising from
decays of heavy resonances of fixed species, even if one neglects
their  life-times, is essentially non-Gaussian, especially at
small pion transverse momenta. Then, if one gets the
interferometry radii from a Gaussian fit to the correlation
function in the region of the interferometry peak, the
longitudinal radii will be larger than the homogeneity lengths of
regions where the spectra of ''resonance
pions" are formed. This violates the natural physical correlation $%
\left\langle f(t,p)\right\rangle \sim
\frac{d^{3}N}{d^{3}p}/V_{int}$  that is typical for direct
particles. Accounting for the finite life-times of resonances
makes the picture even more complicated: the emission is now
spread out in time direction and therefore the $V_{int}$ cannot be
attributed to the thermal freeze-out hypersurface. We have argued
thereby that resonance contributions should be excluded when the
averaged pion phase-space densities at the freeze-out stage are
determined. Our estimates of the total and thermal spectra and the
ratios of correspondent interferometry volumes at different
intensities of transverse flow, including what are typical for SPS
and RHIC energies, gives the possibility of restoring the APSD of
thermal pions at thermal freeze-out. We argue that the values
found are the maximal (total) pion APSD at and after the
decoupling of the hydrodynamic system.

We found that the overpopulation of thermal pions at the thermal
freeze-out, $100 \leq T_{th} \leq 120$ MeV, is associated with the
pion chemical potentials $15 - 51$ MeV for Pb+Pb(Au) 158 AGeV
collisions at  CERN  SPS and $42 - 64$ MeV for Au+Au
$\sqrt{s_{NN}}=130$ GeV collisions at BNL RHIC. The estimates of
the correspondent values in the recent paper \cite{freeze-out}
based on the concrete hydrodynamic parametrization are close to
our results for SPS energies, while for RHIC energies the pions
chemical potentials estimated in Ref. \cite{freeze-out} are
somewhat higher. The thermal pion densities turn out to be
essentially smaller than the critical ones. The correspondent
values  at the SPS energies are $0.014 - 0.017<0.065$ fm$^{-3}$
for $T_{th}=100$ MeV and $0.022 - 0.027<0.095$ fm$^{-3}$ for
$T_{th}=120$ MeV. For the RHIC energies we found $0.018 -
0.020<0.065$ fm$^{-3}$ for $T_{th}=100$ MeV and $0.028 -
0.031<0.095$ fm$^{-3}$ for $T_{th}=120$ MeV. Thereby the
multibosonic effects at those energies would be considered rather
as a correction factor than as an important physical phenomenon.

It looks  puzzling that the averaged phase- space densities of
thermal pions at RHIC are  higher than at SPS, $0.144 -
0.168>0.105 - 0.137$ while the particle freeze-out densities are
comparable. The detailed analysis of the evolution of the APSD
could explain such a behaviour. Indeed, if one imagines   that at
some initial time hadronic system  has no significant transverse
flow, then the averaged PSD is proportional to the PSD  and
particle density,  as in a hot, dense gas in a box, and  all the
densities are high. However, when the system expands and
transverse flow develops, the behaviors of the APSD and particle
density are completely different. While the pion density is
reduced dramatically, at the last stage as $t^{-3}$, the pion APSD
changes (it can even increase) rather slowly. Thereby the systems
reach kinetic freeze-out at some typical particle densities that
are similar for different collision energies,  and, at the same
time, their  APSD could increase with energy  since  it
"memorizes" the higher initial hadronic densities at higher
energies. It is found that such a "memory" is conditioned by the
conservation of entropy and particle numbers during chemically
frozen hydrodynamic evolution of the mixture of hadronic gases.

The main aspect of the HBT puzzle concerns the energy dependence
of the interferometry volume that turns out essentially smaller
than  expected at the RHIC energies. Our results could shed light
on this phenomenon. Since the pion APSD and effective transverse
temperature do not change essentially during the evolution, the
''interferometry volume'', $V_{int}\sim
(dN/dy)/(T_{eff}^{3}\left\langle f(t)\right\rangle )$, defined
formally at any time during the evolution of hadronic system, also
changes rather slowly with time. The concrete mechanism of the
effect is that, if the intensity of the transverse flow grows
during the evolution, the contribution of an enlarged geometric
radius of the expanding system to the interferometry volume is
almost compensated by a stronger reduction of the  interferometry
radii due to the strengthening of the flow (see, e.g.,
\cite{AkkSin}). The initial effective volume is, thus, partially
frozen in the form of the interferometry volume. There is a rough
similarity  between  the effective,  or "interferometry" volumes
at higher (RHIC) and smaller (SPS) energies because the larger
configuration volume at hadronization stage at RHIC (particle
densities at the hadronization are approximately the same as at
the SPS) is partially compensated by the higher intensities of
flow. Therefore, the unexpectedly small increase of the
interferometry volume at  RHIC is caused by strengthening of
transverse flows at higher energies. During the evolution of
hadronic system  the APSD is reduced rather slowly (''frozen'')
while the particle density falls down quickly until it reaches the
standard conditions for system decay. That is the reason why the
observed value of the APSD has no direct link to the freeze-out
criteria and final thermodynamic parameters, being connected
rather to the initial phase-space density of hadronic matter
formed in relativistic nucleus-nucleus collisions.

\section*{Acknowledgments}

We are grateful to P. Braun-Munzinger for his interest in this
work and stimulating discussions, and S. Steinke for careful
reading of the manuscript and useful suggestions. The work was
supported by NATO Collaborative Linkage Grant No. PST.CLG.980086,
Ukrainian State Fund of the Fundamental Researches, Project No.
2.7/135, and US Civilian Research and Development Foundation
(CRDF) Cooperative Grants Program,   Project Agreement UKP1-2613-KV-04. 
Research carried out within the scope of the ERG (GDRE): Heavy ions at
ultrarelativistic energies -- a European Research Group comprising
IN2P3/CNRS, Ecole des Mines de Nantes, Universite de Nantes,
Warsaw University of Technology, JINR Dubna, ITEP Moscow and
Bogolyubov Institute for Theoretical Physics NAS of Ukraine.

\newpage

\begin{figure}[h]
\centering
\includegraphics[scale=0.5]{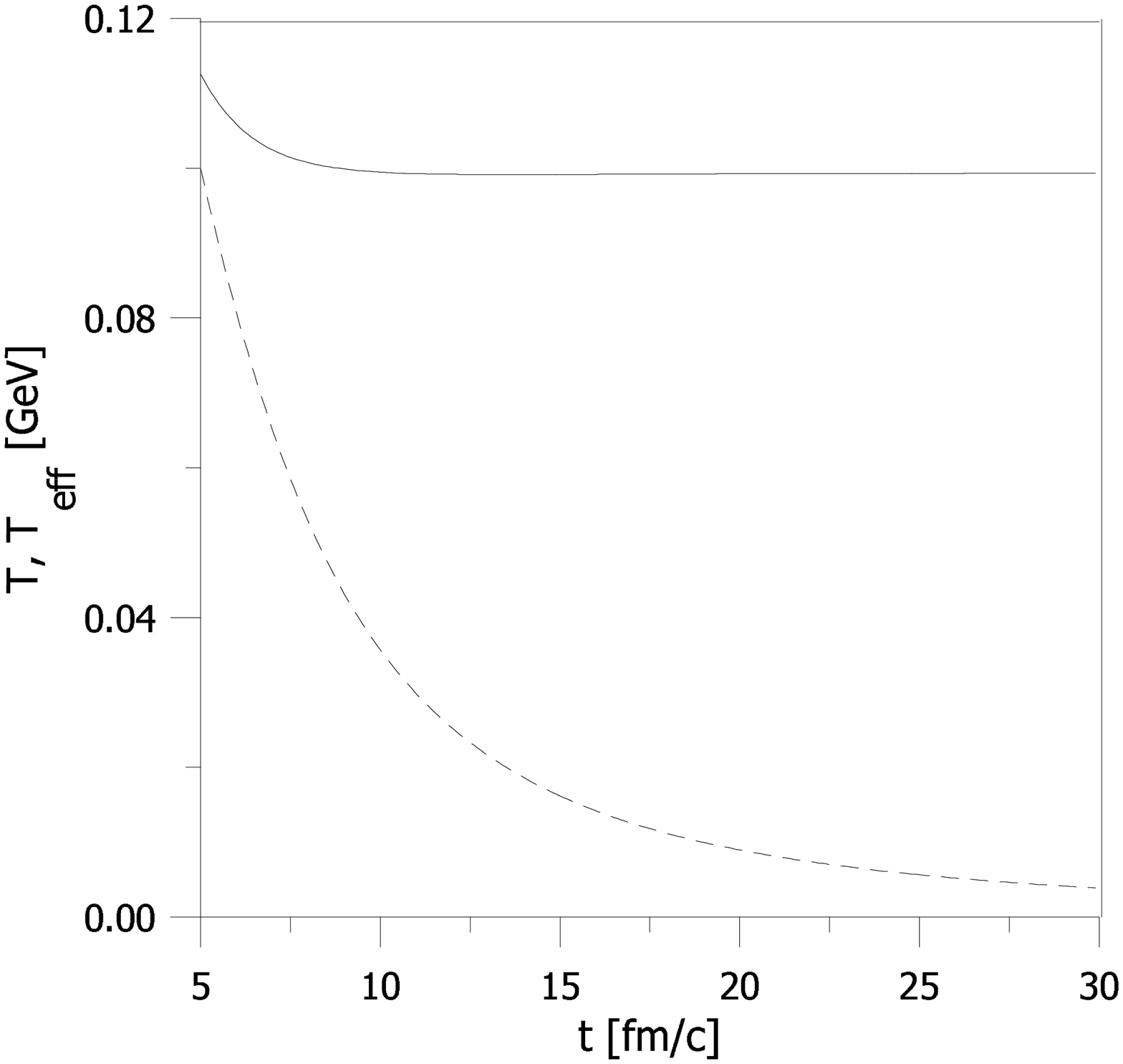}
\caption{ Evolution of the pion effective temperature $T_{eff}$
(solid line) versus the temperature of the system $T$
 (dashed line) in the non-relativistic one-component
Bjorken-like model with transverse expansion. The plots correspond
 to initial conditions: $T_{0}=0.1$  GeV, time $t_{0}=5$ fm/c, transverse radius $%
R_{0}=5$ fm and initial transverse flow $\stackrel{\cdot }{R}%
(t_{0})=0.3$.} \label{fig1}
\end{figure}

\newpage

\begin{figure}[h]
\centering
\includegraphics[scale=0.5]{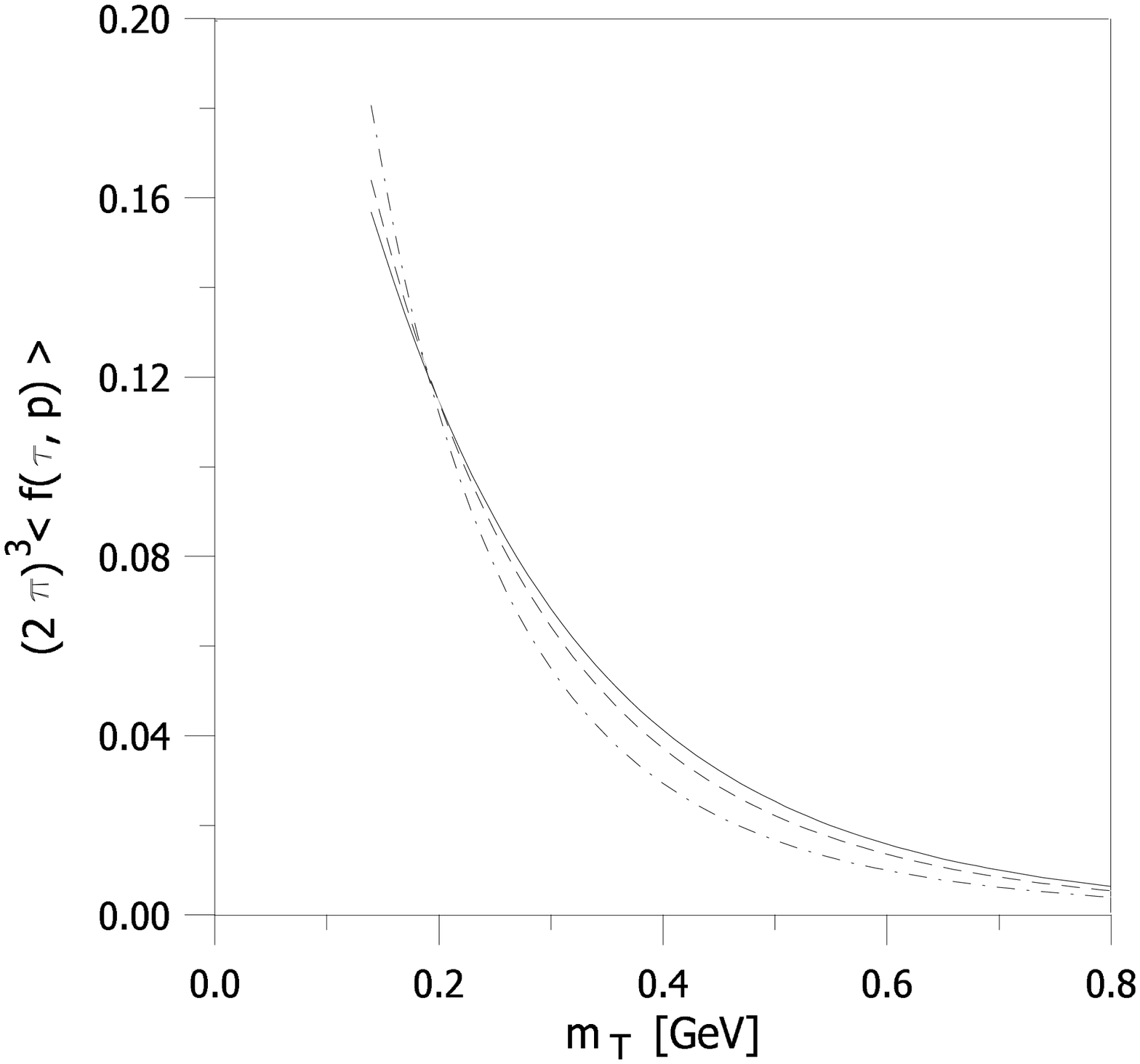}
\caption{ The average phase-space densities $\left\langle f(\tau
,p)\right\rangle $ at typical proper times: at hadronization,
$\tau =7.24$ fm/c, (solid line) and at kinetic freeze-out $\tau
=8.9$ fm/c (dashed line). The dot-dashed line corresponds to the
''asymptotic'' time $\tau =15$ fm/c of hydrodynamic solutions
continued beyond kinetic freeze out. The initial conditions of
hydrodynamic expansion are taken from Table 1 and based on Ref.
\cite{A-B-S}.} \label{fig2}
\end{figure}

\newpage

\begin{figure}[h]
\centering
\includegraphics[scale=0.5]{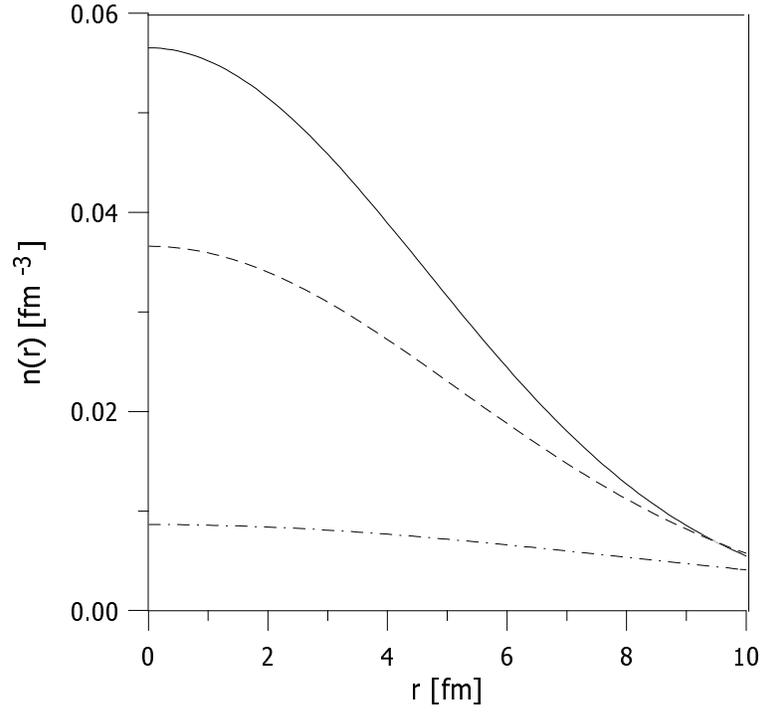}
\caption{ The space densities $n(r)$  taken at typical proper
times: at hadronization, $\tau =7.24$ fm/c, (solid line) and at
kinetic freeze-out $\tau =8.9$ fm/c (dashed line). The dot-dashed
line corresponds to the ''asymptotic'' time $\tau =15$ fm/c of
hydrodynamic evolution continued beyond kinetic freeze out. The
initial conditions of the hydrodynamic expansion are taken from
Table 1 and based on Ref. \cite{A-B-S}. } \label{fig3}
\end{figure}

\newpage

\begin{figure}[h]
\centering
\includegraphics[scale=0.5]{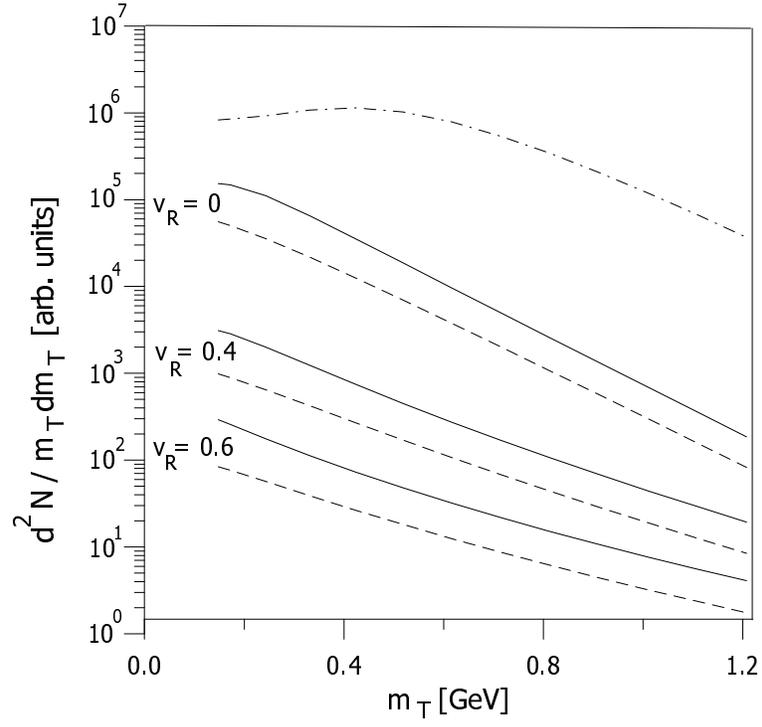}
\caption{The pion transverse spectra for the complete pion yields
(solid lines) and for ''direct'' pions only (dashed line) in the
cases with and without transverse flow: $v_R=0$, $0.4$  and $0.6$.
The shape of the pion spectra from the fixed heavy resonance
species $f_{2}(1270)$ is presented for illustration by dot-dashed
line for $v_R=0$. The overall normalization is arbitrary.}
\label{fig4}
\end{figure}

\newpage

\begin{figure}[h]
\centering
\includegraphics[scale=0.5]{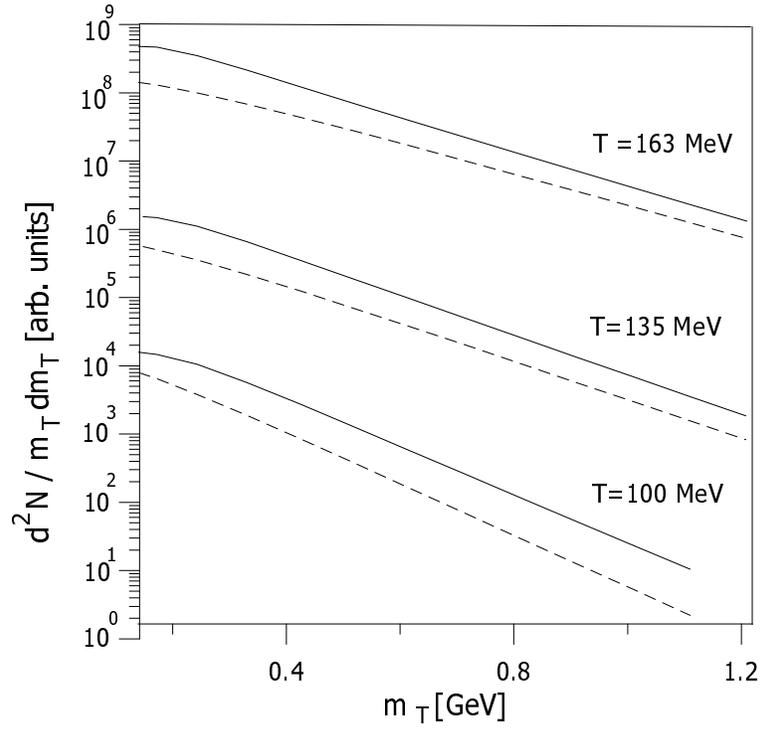}
\caption{The pion transverse spectra for the complete pion yields
(solid lines) and for ''direct'' pions only (dashed line)  taken
at a few typical temperatures
 without transverse flow, $v_R=0$. The overall normalization is arbitrary.}
\label{fig5}
\end{figure}

\newpage

\begin{figure}[h]
\centering
\includegraphics[scale=0.5]{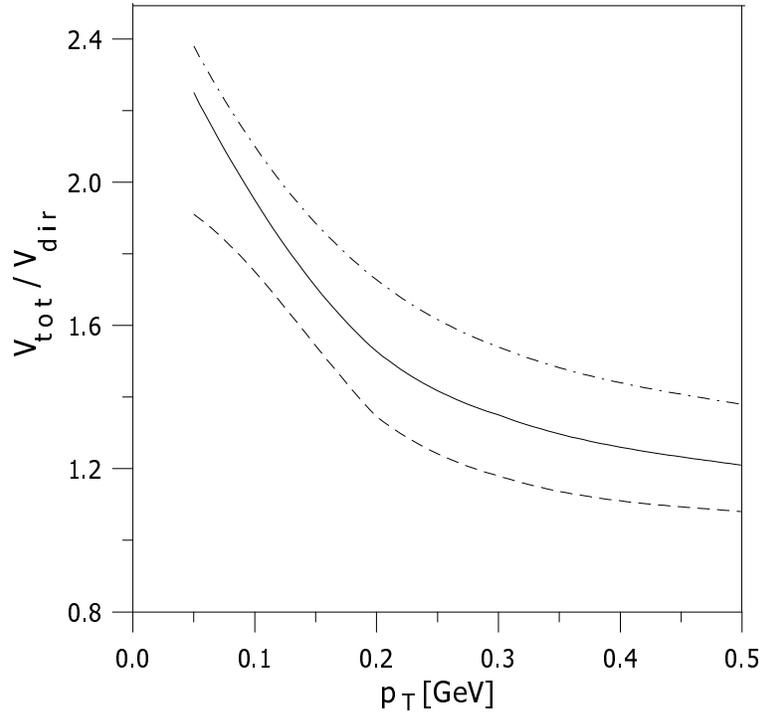}
\caption{ The ratio of the complete interferometry volume
$V_{tot}$, that includes effects of the resonance decays to pions,
to the interferometry volume for ''direct'' pions taken at a few
typical intensities of flow. The resonance contributions are
calculated in accordance with the chemical freeze-out conception,
when approximately $2/3$ of pions are produced by resonance
decays. The dashed line corresponds to $v_{R}=0$, solid line:
$v_{R}=0.4$  and dash-dotted line: $v_{R}=0.6 $. The other
parameters are taken from Ref. \cite{A-B-S} as  described in the
text. } \label{fig6}
\end{figure}

\newpage

\begin{figure}[h]
\centering
\includegraphics[scale=0.5]{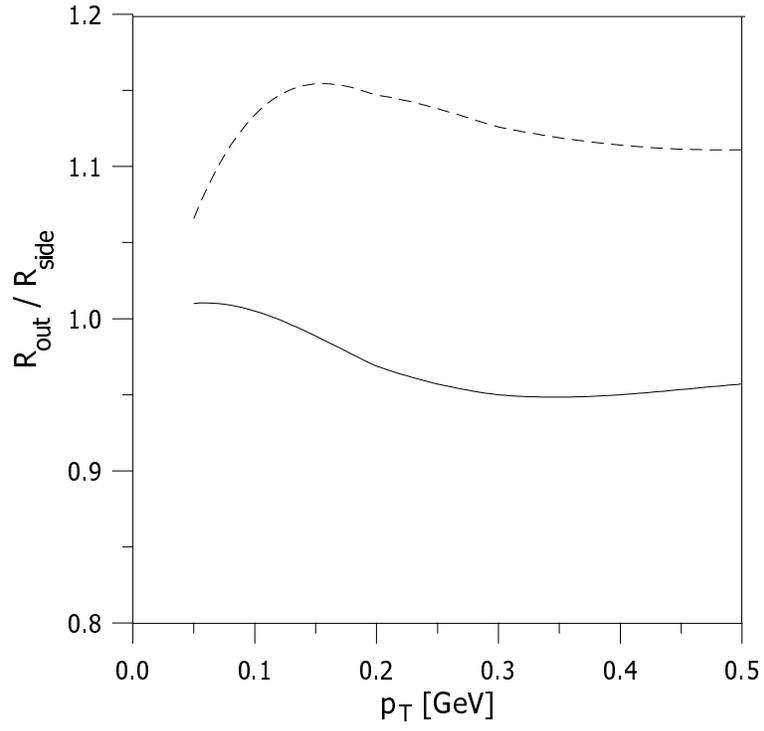}
\caption{ The $R_{out}$ to $R_{side}$ ratio for complete
interferometry radii (solid line) vs ones for ''direct'' pions
(dashed line) at transverse flow $v_{R}=0.6$ which is associated
with RHIC energies. } \label{fig7}
\end{figure}

\newpage

\begin{figure}[h]
\centering
\includegraphics[scale=1.0]{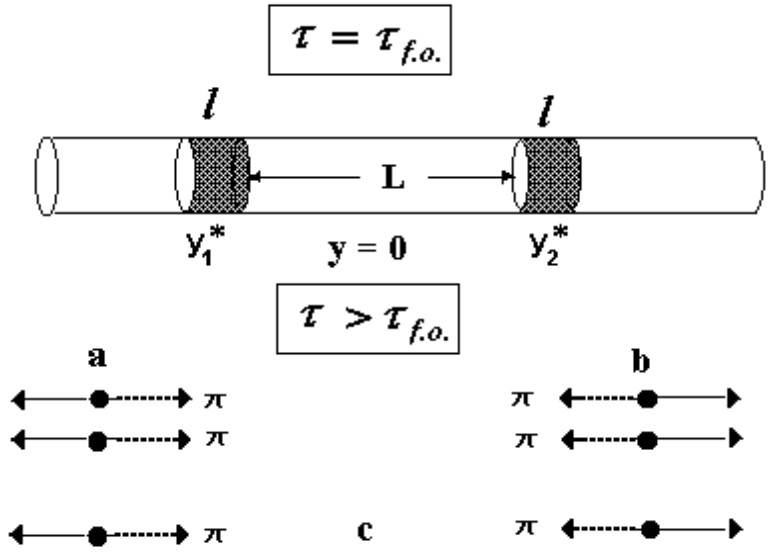}
\caption{The two- pion spectra formation at $p_{T}\simeq 0$ and
$y\simeq 0$ due to decays of heavy resonances involved in the
longitudinal flow before and  at the  freeze-out time
$\tau_{f.o.}$, see the text for details. } \label{fig8}
\end{figure}

\newpage

\begin{figure}[h]
\centering
\includegraphics[scale=1.0]{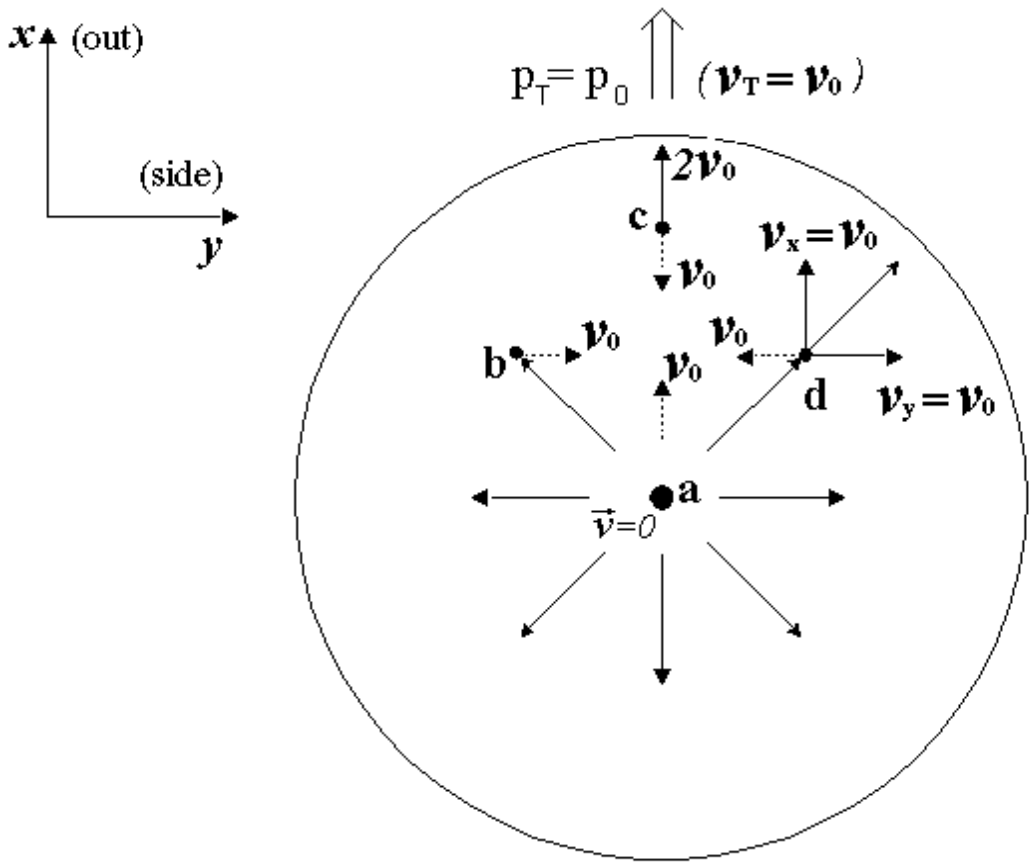}
\caption{The formation of the pion transverse
 spectra for decays of the heavy resonances involved in transverse flows
 as it is  described  in the text.  The solid arrows are associated with radial  flow,
 the  dashed arrows correspond to emission of pions from the resonances.} \label{fig9}
\end{figure}

\end{document}